\documentclass[useAMS]{mn2e}
\usepackage{psfig}

\usepackage{times}
\def\spose#1{\hbox to 0pt{#1\hss}}
\newcommand\lsim{\mathrel{\spose{\lower 3pt\hbox{$\mathchar"218$}}
     \raise 2.0pt\hbox{$\mathchar"13C$}}}
\newcommand\gsim{\mathrel{\spose{\lower 3pt\hbox{$\mathchar"218$}}
     \raise 2.0pt\hbox{$\mathchar"13E$}}}
\def\ltsima{$\; \buildrel < \over \sim \;$}
\def\lsim{\lower.5ex\hbox{\ltsima}}
\def\gtsima{$\; \buildrel > \over \sim \;$}
\def\gsim{\lower.5ex\hbox{\gtsima}}

\def\sch{Schwarzschild}
\title[High--z blazars with GROND and Swift]
{High redshift {\it Fermi} blazars observed by GROND and {\it Swift}}

\author[Ghisellini et al.] 
{G. Ghisellini$^1$\thanks{E--mail:gabriele.ghisellini@brera.inaf.it}, 
M. Nardini$^2$, Tagliaferri$^1$, J. Greiner$^3$, P. Schady$^3$, A. Rau$^3$,
\newauthor
L. Foschini$^1$, F. Tavecchio$^1$, G. Ghirlanda$^1$, T. Sbarrato$^{1,4}$ \\
$^1$: INAF -- Osservatorio Astronomico di Brera, via E. Bianchi 46, I--23807 Merate, Italy \\
$^2$ Univ. di Milano Bicocca, Dip. di Fisica G. Occhialini, Piazza della Scienza 3, I--20126 Milano, Italy \\
$^3$ Max Planck Institut f\"ur extraterrestrische Physik, Giessenbachstrasse 1, 85748 Garching, Germany \\
$^4$ Univ. dell'Insubria, Dipartimento di Fisica e Matematica, Via Valleggio 11, I--22100 Como, Italy \\
}

\begin{document}

% \date{Accepted 1988 December 15. Received 1988 December 14; 
% in original form 1988 October 11}

\pagerange{\pageref{firstpage}--\pageref{lastpage}} \pubyear{2007}

\maketitle

\label{firstpage}

\begin{abstract}
We  observed 5 $\gamma$--ray loud blazars at redshift greater than 2 
with the X--Ray Telescope (XRT) and the UltraViolet and Optical Telescope
(UVOT) onboard the {\it Swift} satellite, and  the
Gamma--Ray burst Optical Near--Infrared Detector (GROND) instrument.
These observations were quasi simultaneous, usually within a few hours.
For 4 of these blazars the near--IR to UV data show the presence of
an accretion disc, and we could reliably estimate its accretion rate and black hole mass.
One of them, PKS 1348+007, was found in an extraordinarily high IR--optical state, 
almost two orders of magnitude brighter than at the epoch of the Sloan Digital Sky Survey
observations.
For all the 5 quasars the physical parameters of the jet emitting zone, 
derived by applying a one--zone emission model,
are similar to that found for the bulk of other $\gamma$--ray loud quasars.
With our observations we have X--ray data for the full sample of blazars
at $z>2$ present in the {\it Fermi} 2--yrs (2LAC) catalog.
This allows to have a rather complete view of the spectral energy
distribution of all high--redshift {\it Fermi} blazars, 
and to draw some conclusions about their properties,
and especially about the relation between the accretion rate and the jet power.
\end{abstract}


\section{Introduction}

High redshift blazars are the most powerful persistent
sources, and are usually connected with the most massive black holes 
(Ghisellini et al. 2009; 2010a; Volonteri et al. 2011).
% The most important results have been achieved by observations of 
The Large Area Telescope (LAT) onboard the {\it Fermi} satellite (Atwood et al. 2009)
and the Burst Alert Telescope (BAT) onboard {\it Swift} (Gehrels et al. 2004) 
have provided tens of detections 
of blazars at $z>2$ (Abdo et al. 2010, Ackermann et al. 2011; Ajello et al. 2009, 
Cusumano et al. 2010; Baumgartner et al. 
2010\footnote{online data in http://heasarc.gsfc.nasa.gov/docs/swift/results/bs58mon/}).
The combinations of the two sets of data ({\it Swift+Fermi}) have allowed to measure
the properties and the bolometric luminosity of the jet non--thermal emission 
and to characterize the thermal component emitted by 
the accretion disc, namely the black hole mass $M$ and the accretion rate $\dot M$.
It is found (Ghisellini et al. 2010b; 2011)
that in powerful blazars black hole masses are usually greater than $10^9 M_\odot$,
with disc luminosities $L_{\rm d}\sim$0.1$L_{\rm Edd}$,
jet kinetic powers $P_{\rm j} \sim \dot M c^2$, with
$P_{\rm j}$ tightly related to $L_{\rm d}$.
This can test jet production models, since $P_{\rm j}\gsim \dot M c^2$
requires that we are using another source of energy besides accretion, namely
we must extract the black hole spin energy (see e.g. Tchekhovskoy, Narayan \& McKinney 2011).

In order to find the physical parameters of these 
sources it is very important to have a good coverage of their spectral 
energy distribution (SED), that allows us to constrain the model parameters.
Moreover, blazars are varying fast (hours to days, especially at high
frequencies), and with large amplitudes
(even by factor 10 or more in the $\gamma$--ray range, but sometimes even
in the optical) implying that simultaneous observations are needed to well
constrain their SED.

High redshift blazars are interesting ``per se", since we would like to know 
if and how the jet properties change with cosmic time.
High redshift also usually means larger luminosities and powers, 
and this allows to study the more powerful jets.
If we {\it define} a blazar
as a source whose jet is observed with a viewing angle $\theta_{\rm v}<1/\Gamma$
($\Gamma$ is the bulk Lorentz factor),
then there must be other $2\Gamma^2$ similar objects whose jet
is pointing elsewhere and whose flux is dramatically fainter
(because of beaming).
These misaligned sources share all the intrinsic properties
of the blazar that is pointing at us, including the black hole mass.
If we are able to estimate it for a blazar, we
can put very interesting constraints on the density of heavy black holes
of all radio sources in the young Universe (see Volonteri et al. 2011).

From the $z>2$ blazars detected by {\it Fermi}/LAT, we have chosen
those with no data in the X--ray range (or with only an upper limit),
and have organized a simultaneous observational campaign
involving the X--Ray Telescope (XRT) and the UltraViolet and Optical Telescope
(UVOT) onboard {\it Swift} (Gehrels et al. 2004) and  the
Gamma--Ray burst Optical Near--Infrared Detector (GROND) instrument 
(Greiner et al. 2008).
The scientific rationale for observing these blazars at X--ray 
and IR--optical frequencies is the following.
In powerful blazars, the [0.3--10 keV] {\it Swift} band is where
we expect the contribution of the inverse Compton emission of the jet.
This is usually made by two components, according if the scattering
process makes use of internally produced synchrotron seed photons
(Synchrotron Self Compton, or SSC for short) or if the seeds
are produced externally to the jet (External Compton, EC).
They are usually characterized by a different spectrum 
and variability behavior, and
the relative importance of the two gives information on
the magnetic field and the bulk Lorentz factor.
In the near IR, optical and UV bands, instead, we have the 
contributions of the accretion disc and the beamed synchrotron
component. 
If we can distinguish the two contributions then we can
estimate the black hole mass and the accretion rate (from the disc emission),
and have information of the value of the magnetic field
(from the synchrotron flux).

The starting samples were the {\it Fermi}/LAT detected blazars at $z>2$
present in the {\it Fermi} blazar catalogs after 11 months of operations 
(1LAC, Abdo et al. 2010) and after 2 years (2LAC, Ackermann et al. 2011).
In the 1LAC catalog there are 2 sources observed by {\it Swift} 
in 2009 for just a ks, for which we could only derive an upper limit 
to the X--ray flux.
In the 2LAC catalog there are 3 new blazars at $z>2$ (out of 31)
without a proper characterization of their SED, because of no information
on their X--ray flux, useful to characterize the jet beamed
emission, nor good data coverage in the IR--optical--UV band,
useful to derive the thermal (i.e. accretion disc) contribution.
These 5 blazars are all at declination $<$+30$^\circ$, and are visible from
La Silla, where the GROND instrument operates.
Therefore we organized quasi--simultaneous {\it Swift} and GROND observing 
campaigns for these blazars. 

These observations provided us with optical--UV--X--ray information on
the entire high redshift {\it Fermi} sample (``clean" 2LAC with $z>2$).
Tab. \ref{sample} lists the 5 selected blazars, together with their $\gamma$--ray 
k--corrected luminosities.
These are the averaged luminosity over 11 months (1LAC) and 2 yrs (2LAC). For 
the blazars 1149--084 and 1344--1723, present in both catalogs, we give
the corresponding two luminosities.

With a good simultaneous coverage from the near IR to X--ray range, in addition to
the {\it Fermi} data,
we can properly characterize the SED in order to disentangle the non--thermal jet
contribution and the thermal component, to find the black hole mass, the accretion rate,
and the physical jet quantities (magnetic field, bulk Lorentz factor, particle densities and
jet power).

We use a flat cosmology with $H_0 =70$ km s$^{-1}$ Mpc$^{-1}$,  $\Omega_{\rm M}$=0.3
and the notation $Q=10^X Q_X$ in cgs units.

%---------------------------------
\begin{table} 
\centering
\begin{tabular}{llllll}
\hline
\hline
Name       &RA          &Dec  &$z$ &$L_{\gamma,1}$  &$L_{\gamma,2}$  \\% L_BLR =4 Lgamma^0.93
\hline   
PKS 0519+01      &05 22 17.5  &+01 13 31    &2.941 &...  &47.0  \\  % 44.31
TXS 1149--084    &11 52 17.2  &--08 41 03   &2.367 &47.7 &47.7  \\  % 44.96
MG2 J133305+2725 &13 33 07.5 &+27 25 18    &2.126 &...  &47.3  \\  % 44.59
PMN J1344--1723  &13 44 14.4  &--17 23 40   &2.49  &48.5 &48.2  \\  % 45.44 
PKS 1348+007     &13 51 04.4  &+00 31 19    &2.084 &...  &47.6  \\  % 44.87
\hline
\hline 
\end{tabular}
\vskip 0.4 true cm
\caption{List of our sources.
$L_{\gamma, 1}$ refers to the [0.1--10 GeV] $\gamma$--ray luminosity in the 1LAC catalog,
while $L_{\gamma, 2}$ is the one in the 2LAC catalog, in units of erg s$^{-1}$.
} 
\label{sample}
\end{table}
%---------------------------------

\section{GROND observations and data analysis}

The 7--band  GROND imager is mounted at the 2.2 m MPG/ESO telescope 
at La Silla Observatory (Chile). 
GROND is able to observe {\it simultaneously} in 7 filters,
from the NIR $Ks$ (2300 nm) to the $g^\prime$ band (360 nm).
Therefore it nicely complements the UVOT filters, with the bluest ($g^\prime$ and $r^\prime$)
filter overlapping in part with the reddest $v$ and $b$ UVOT filter (and this
is useful for cross calibration).

We carried out observations for all sources simultaneously in all 
7 $g^\prime, r^\prime, i^\prime, z^\prime, J, H, K_{\rm s}$ bands. 
The log of the GROND observations and the related observing conditions are reported 
in Tab. \ref{grondlog}. 

The GROND optical and NIR image reduction and photometry were
performed using standard IRAF tasks (Tody 1993), similar to the
procedure described  in Kr\"uhler et al. (2008). 
A general model for the point--spread function (PSF) of each image was
constructed using bright field stars, and it was then fitted to the point source. 
When the source field was covered by the SDSS (Smith et al. 2002) survey 
(i.e. PKS 0519+01, PKS 1348+007, MG2 J133305+2725), the absolute calibration of 
the $g^\prime, r^\prime, i^\prime, z^\prime$ bands was  obtained with respect to 
the magnitudes of SDSS stars within the blazar field. 
In the other cases (i.e. PMN J1344--1723 and TXS 1149--084), optical 
photometric calibration was performed relative to the magnitudes of 
six secondary standards in the blazar field. 
During photometric conditions, a primary SDSS standard
field  was observed within minutes of an observation of the source field.
The obtained zero--points were corrected for atmospheric extinction and used to 
calibrate stars in the blazar field. 
The apparent magnitudes of the sources were measured with respect to 
these secondary standards. For all sources the  $J, H, K_{\rm s}$ 
bands calibrations were obtained with respect to magnitudes of 
the Two Micron All Sky Survey (2MASS) stars (Skrutskie et al. 2006).

Tab. \ref{grond} reports the observed AB magnitudes, not corrected for the 
Galactic extinction listed in the last column and taken from 
Schlafly \& Finkbeiner (2011).

\section{{\it Swift} observations and data analysis}
 
We have analysed the {\it Swift} X--Ray Telescope (XRT; Burrows et al. 2005) 
and Optical--Ultraviolet Telescope (UVOT; Roming et al. 2005) data.
The data were screened, cleaned and analysed with the software package
HEASOFT v. 6.12, with the calibration database updated to 22 March 2012.
The XRT data were processed with the standard procedures ({\texttt{XRTPIPELINE v.0.12.6}). 
All sources were observed in photon counting (PC) mode and grade 0--12 
(single to quadruple pixel) were selected. 
The channels with energies below 0.3 keV and above 10 keV were excluded from the fit 
and the spectra were rebinned in energy so to have at least 20--30 counts per bin
in order to apply the $\chi^2$ test. 
When there are no sufficient counts
we applied the likelihood statistic in the form reported by Cash (1979).
Each spectrum was analysed in XSPEC v. 12.7.1
with an absorbed power law model with a fixed Galactic column density as
measured by Kalberla et al. (2005). 
The computed errors represent the 90\% confidence interval on the spectral parameters. 
Tab. \ref{xrt} 
reports the log of the observations and the best fit results of 
the X--ray data with a simple power law model. 
The X--ray spectra displayed in the SED have been rebinned to ensure the best 
visualization.

UVOT source counts were extracted from a circular region 5"--sized centred 
on the source position, while the background was extracted from an 
annulus with internal radius of 7" and variable outer radius 
depending on the nearest contaminating source.
% larger circular nearby source--free region.
Data were integrated with the \texttt{uvotimsum} task and then 
analysed by using the  \texttt{uvotsource} task.
The observed magnitudes have been dereddened according to the formulae 
by Cardelli et al. (1989) and converted into fluxes by using standard 
formulae and zero points.
%  from Poole et al. (2008).
% Tab. \ref{uvot} lists the observed magnitudes.
% and the Galactic extinction appropriate for each source.
%---------------------------------

As can be seen, the UVOT observations yielded mostly upper limits,
and very few detections. 
The listed UVOT and GROND magnitudes do not take into account any difference
between the two instruments, that is however likely, and of the order of 0.1--0.3
magnitudes (see Rau et al. 2012), especially if the observations were not
exactly simultaneous, but separated by a few hours (the maximum separation -- two days --
occurs for PKS 0519+01).

% ------------------------------------------------------------------------
\begin{table*} 
\centering
\begin{tabular}{llllll }
\hline
\hline
Name       &Date         &Start time     &Exp: opt/IR &Average seeing  &Average airmass \\
           &yyyy--mm--dd &$[{\rm UTC}]$  &[s]      &[arcsec]        &                \\
\hline   
0519+01    &2012--02--26  &00:14:00      &919/960     &0.70            &1.18  \\  
1149--084  &2012--03--20  &02:09:51      &1501/1200   &0.72            &1.27  \\
1333+2725  &2012--05--09  &02:40:11      &426/720     &1.31            &1.83  \\
1344--1723 &2012--04--04  &08:36:13      &919/960     &1.32            &1.44  \\
1348+007   &2012--03--16  &05:03:37      &3002/2400   &1.15            &1.26  \\
1348+007   &2012--03--18  &09:38:11      &460/480     &1.43            &1.61  \\
\hline
\hline 
\end{tabular}
\vskip 0.4 true cm
\caption{
Log of the GROND observations. 
Exposures refer to optical/NIR filters while the average seeing is calculated in 
the $r^\prime$ band.
} 
\label{grondlog}
\end{table*}
% ------------------------------------------------------------------------

% ------------------------------------------------------------------------
\begin{table*} 
\centering
\begin{tabular}{lllll llll }
\hline
\hline
Name    &$g'$ &$r'$ &$i'$ &$z'$ &$J$ &$H$ &$K_s$  &$A_V$\\
\hline   
0519+01       &20.37$\pm$0.05 &19.83$\pm$0.05 &19.50$\pm$0.05 &19.28$\pm$0.05 &18.74$\pm$0.11 &18.54$\pm$0.11 &18.61$\pm$0.15 &0.38  \\  
1149--084     &19.65$\pm$0.05 &19.42$\pm$0.06 &19.40$\pm$0.08 &19.04$\pm$0.08 &18.97$\pm$0.11 &18.67$\pm$0.12 &18.24$\pm$0.13 &0.23  \\
1333+2725     &20.38$\pm$0.06 &19.89$\pm$0.06 &19.39$\pm$0.06 &19.07$\pm$0.05 &18.33$\pm$0.10 &17.73$\pm$0.10 &17.44$\pm$0.13 &0.03   \\
1344--1723    &20.69$\pm$0.05 &20.26$\pm$0.05 &19.99$\pm$0.05 &19.56$\pm$0.06 &19.01$\pm$0.12 &18.42$\pm$0.12 &18.11$\pm$0.13 &0.37  \\
1348+007 (1)  &19.65$\pm$0.05 &19.25$\pm$0.05 &18.90$\pm$0.05 &18.62$\pm$0.06 &18.25$\pm$0.10 &17.79$\pm$0.10 &17.40$\pm$0.12 &0.11  \\
1348+007 (2)  &20.24$\pm$0.05 &19.88$\pm$0.05 &19.49$\pm$0.05 &19.19$\pm$0.06 &18.66$\pm$0.11 &18.20$\pm$0.11 &17.76$\pm$0.12 &0.11  \\
\hline
\hline 
\end{tabular}
\vskip 0.4 true cm
\caption{Observed magnitudes, not corrected for Galactic foreground reddening, in the AB system.
Errors include systematics.
%{\bf
%I increased the errors by adding 0.05 for optical and 0.1 in Infrared...
%}
The last column reports the value of the Galactic $A_V$ from Schlafly \& Finkbeiner (2011).
}
\label{grond}
\end{table*}
% ------------------------------------------------------------------------

%---------------------------------
\begin{table*} 
\centering
\begin{tabular}{llllllllllllll}
\hline
\hline
Name   &Date              &Exp. &$N^{\rm Gal}_{\rm H}$ &$\Gamma$ &$F_X$        &$\chi^2$ (d.o.f.) &Note\\
       &yyyy--mm--dd/UTC) &ks   & cm$^{-2}$            &         &$10^{-13}$ cgs \\
\hline   
0519+01    &2012--02--24/01:26                        &21.9 &1.07e21 &1.76$\pm$0.24 &3.1  &76.28 (95) &Cash stat \\ % &0.11 (2) & \\  
1149--084  &2012--03--19/00:10 $+$ 2012--03--20/08:09 &17.1 &4.75e20 &1.49$\pm$0.28 &2.5  &70.02 (67) &Cash stat \\ % &0.01 (1) & \\
1333+2725  &2012--05--09/00:09 $+$ 2012--05--10/21:16 &19.8 &1.02e20 &1.84$\pm$0.17 &6.65 &10.87 (12) &   \\ 
1344--1723 &2012--04--03/10:55 $+$ 2012--04--04/01:41 &14.9 &8.70e20 &1.87$\pm$0.40 &1.5  &36.68 (39) &Cash stat \\
1348+007   &2012--03--16/04:51 $+$ 2012--03--18/13:17 &17.5 &2.22e20 &1.50$\pm$0.26 &2.9  &82.27 (78) &Cash stat \\ % &0.03 (1) & \\
\hline
\hline 
\end{tabular}
\vskip 0.4 true cm
\caption{
Observation log and results of the X--ray analysis of XRT data.
The flux $F_X$ is in the [0.3--10 keV] band, it is de--absorbed with the indicated $N^{\rm Gal}_{\rm H}$
and it is in units of $10^{-13}$ erg cm$^{-2}$ s$^{-1}$. Apart from 0519+01, all other sources
were observed in two occasions: the listed exposure time is the sum. The analysis has been performed 
on the total.
All sources but 1333+2725 have been
analyzed with the Cash statistics (Cash 1979; Gehrels 1986).
} 
\label{xrt}
\end{table*}
%---------------------------------

% ------------------------------------------------------------------------
\begin{table*}  
\centering
\begin{tabular}{lllllll}
\hline
\hline
Name   &$v$ &$b$ &$u$ &$uvw1$ &$uvm2$  &$uvw2$  \\
\hline   
0519+01     &$>$19.1 &$>$20.0 &$>$19.5      &$>$19.6      &$>$21.7      &$>$21.9   \\   
1149--084   &$>$19.1 &$>$20.0 &$>$19.7      &$>$20.0      &$>$20.6      &$>$21.8   \\
1333+2725   &$>$19.2 &$>$20.2 &20.3$\pm$0.3 &20.4$\pm$0.4 &19.7$\pm$0.2 &19.9$\pm$0.1   \\
1344--1723  &$>$19.0 &$>$20.0 &21.0$\pm$0.2 &$>$19.9      &$>$20.1      &$>$20.7  \\
1348+007    &$>$19.1 &$>$20.2 &$>$20.8      &$>$21.0      &20.3$\pm$0.1 &$>$20.8   \\
\hline
\hline 
\end{tabular}
\vskip 0.4 true cm
\caption{UVOT magnitudes. The magnitude lower limits are at the 3$\sigma$ level.
} 
\label{uvot}
\end{table*}
%---------------------------------

\section{Spectral Energy Distributions and modelling}

In Fig. \ref{0519}--\ref{1348} we show the spectral energy
distribution (SED) of our sources.
We complement our near--simultaneous data with archival data
taken from NED and ASDC\footnote{\tt http://tools.asdc.asi.it/SED/}.
We show the {\it Fermi}/LAT data of the 1LAC or 2LAC catalogs,
but also the average flux corresponding to one month of {\it Fermi}
observation, starting 2 weeks before and ending
two weeks after the GROND+{\it Swift} observing time. 

The adopted model is a one--zone and leptonic model, fully
described in Ghisellini \& Tavecchio (2009).
The main properties are summarized in the Appendix, mainly to explain the 
meaning of the parameters listed in Tab. \ref{para} and Tab. \ref{powers}.

In brief, the model assumes that the bulk of the jet 
dissipation takes place
in one zone located at some distance $R_{\rm diss}$ from the black hole.
For simplicity, the emitting region is assumed spherical with a radius $R=\psi R_{\rm diss}$,
with $\psi=0.1$.  
The region is moving with a bulk Lorentz factor $\Gamma$, and is observed under a viewing
angle $\theta_{\rm v}$.
Energetic electrons are injected throughout the source for a time equal to the light crossing 
time $R/c$, and the particle distribution is calculated (through the continuity equation)
at this time, considering radiative losses
and possible electron--positron pair production and their reprocessing.
In the following we briefly discuss the guidelines for the
choice of the main parameters needed for the model.

% --------------------------------------------------
\begin{table*} 
\centering
\begin{tabular}{lllllllllllllll}
\hline
\hline
Name   &$z$ &$R_{\rm diss}$ &$M$ &$R_{\rm BLR}$ &$P^\prime_{\rm i}$ &$L_{\rm d}$ &$B$ &$\Gamma$ 
     &$\gamma_{\rm b}$ &$\gamma_{\rm max}$ &$s_1$  &$s_2$   \\ % &$\gamma_{\rm peak}$ &$U^\prime$\\ % &$\gamma_{\rm c}$
~[1]      &[2] &[3] &[4] &[5] &[6] &[7] &[8] &[9] &[10] &[11]  &[12] &[13] \\ %  &[14]   &[15]   \\
\hline   
0519+01    &2.941 &540 (400) &4.5e9 &486 &0.011 &23.6 (0.035) &1.9  &11    &50  &3e3   &1   &2.5 \\ % &50  &1.4    \\  
1149--084  &2.367 &180 (400) &1.5e9   &561 &0.01  &31.5 (0.14)  &7.9 &15    &300 &800   &--1 &3.3  \\ % &233  &8.7 \\ 
           &      &{\it 720 (600)} &{\it 4e9}   &{\it 849} &{\it 0.015} &{\it 72    (0.12)} &{\it 1.39 }
    &{\it 14}    &{\it 300} &{\it 3e3}   &{\it --1} &{\it 2}   \\ % &{\it 254} &{\it 4.84}  \\  
1333+2725  &2.126 &45 (1500) &1e8   &300 &0.01  &9 (0.6) &5.5 &13  &600 &1.8e3 &0 &2.4 \\ % &300 &5.2   \\
1344--1723 &2.49  &225 (500) &1.5e9  &351 &0.021 &12.3 (0.055) &2.3 &13  &1.4e3 &2.3e3 &0  &2.7   \\ % &853 &5.06 \\
           &      &{\it 330 (1100)}  &{\it 1e9} &{\it 274} &{\it0.027 } &{\it 7.5 (0.05)} &{\it 0.89} 
    &{\it 13}  &{\it 1.4e3}  &{\it 8e3} &{\it --1} &{\it 2.5} \\ % &{\it 1456} &{\it 0.846}  \\
1348 high  &2.084 &60 (500)  &4e8   &115 &8.5e--3 &1.3   (0.022) &4.5   &11    &500 &4.7e3  &--1   &2.2 \\ % &637 &4.59  \\
1348 low   &2.084 &96 (800)  &4e8   &115 &2.2e--3 &1.3   (0.022) &1.3   &15    &300 &3e3    &--1   &2.5 \\ % &311 &6.51  \\
\hline
\hline 
\end{tabular}
\vskip 0.4 true cm
\caption{List of parameters used to construct the theoretical SED.
Not all of them are ``input parameters" for the model: $R_{\rm BLR}$
is uniquely determined from $L_{\rm d}$, and the cooling energy $\gamma_{\rm c}$ 
and $U^\prime$ are derived parameters.
Col. [1]: name;
Col. [2]: redshift;
Col. [3]: dissipation radius in units of $10^{15}$ cm and (in parenthesis) in units of \sch\ radii;
Col. [4]: black hole mass in solar masses;
Col. [5]: size of the BLR in units of $10^{15}$ cm;
Col. [6]: power injected in the blob calculated in the comoving frame, in units of $10^{45}$ erg s$^{-1}$; 
Col. [7]: accretion disc luminosity in units of $10^{45}$ erg s$^{-1}$ and (in parenthesis) in units of $L_{\rm Edd}$;
Col. [8]: magnetic field in Gauss;
Col. [9]: bulk Lorentz factor at $R_{\rm diss}$;
Col. [10] and [11]: break and maximum random Lorentz factors of the injected electrons;
Col. [12]: and [13]: slopes of the injected electron distribution [$Q(\gamma)$] below and above $\gamma_{\rm b}$;
% Col. [14]: random Lorentz factors of the electrons radiating at the synchrotron peak;
% Col. [15]: sum of the radiation and magnetic energy density in the comoving frame.
For all sources we assumed a viewing angle $\theta_{\rm v}=3^\circ$.
The total X--ray corona luminosity is assumed to be in the range 10--30 per cent of $L_{\rm d}$.
Its spectral shape is assumed to be always $\propto \nu^{-1} \exp(-h\nu/150~{\rm keV})$.
The parameters in italics refer to the physical quantities found in Ghisellini et al. (2011).
}
\label{para}
\end{table*}
% --------------------------------------------------

\subsection{Guidelines for the choice of the parameters}

% The good near--IR to UV coverage of our blazars allows to 
% find, separately, the thermal (i.e. accretion disc) and jet components,
% at least when the synchrotron emission is not dominant (as in the high state of
% 1348+007).

\vskip 0.3 cm
\noindent
{\it The luminosity of the accretion disc} --- It can be
estimated directly if the disc is visible {\it and} its spectrum peaks in the observed 
frequency range. 
This occurs for PKS 0519+01, for TXS 1149--084 and most likely for PMN J1344--1723,
even if in this latter blazar there is a strong ``contaminating" synchrotron component.
The overall luminosity $L_{\rm d}$ of a standard accretion disc is roughly 
twice its $\nu L(\nu)$ peak, therefore we can directly estimate $L_{\rm d}$
if we see a thermal peak in the SED.
For MG2 J1333+2725 the IR to UV continuum is dominated by the steep tail 
of the synchrotron flux, while in PKS 1348+007 we have a hint of the
contribution from the accretion disc from the photometric data of the SDSS.
If an optical spectrum is available (as in the case of TXS 1149--084 and PMN J1344--1723,
Shaw et al. 2012), we have additional information from the luminosity
of the broad emission lines.
Through the templates of Francis et al. (1991) and/or of Vanden Berk et al (2001)
we can reconstruct, from the luminosity of one or more lines,
the entire luminosity of the broad line region, $L_{\rm BLR}$.
Then, applying a typical covering factor $C$ (namely, $C\sim 0.1$),
we can estimate $L_{\rm d}$.
Finally, if no spectrum is available and the thermal component is
completely swamped by the synchrotron flux, we can have a (rough)
indication of $L_{\rm BLR}$ through the correlation between
$L_{\rm BLR}$ and $L_\gamma$ as found by Sbarrato et al. (2012a).
This has the form
\begin{equation}
L_{\rm BLR} \sim 4 L_\gamma^{0.93} 
\end{equation}
Note, however, that since $L_\gamma$ can vary even by two orders of
magnitude in the same source, the correlation necessarily has a large scatter,
making the estimate of $L_{\rm BLR}$ (and thus of $L_{\rm d}$) 
uncertain.

\vskip 0.3 cm
\noindent
{\it Black hole mass} --- 
If $L_{\rm d}$ is determined reliably, there is only one black hole mass value
that can fit the flux produced by (a standard) accretion disc.
In this case the derived mass is robust, with an associated uncertainty of
less than a factor 2 (see Fig. 1 in Sbarrato et al. 2012b, \S4.4 and Fig. \ref{0519} below).
If $L_{\rm d}$ is uncertain, this reflects also on the uncertainty
on the derived black hole mass.

\vskip 0.3 cm
\noindent
{\it Location of the emitting region} --- One of the specific features of the model
is that it calculates the energy densities (magnetic and radiative) 
as a function of the distance $R_{\rm diss}$ from the black hole.
In particular, if $R_{\rm diss}<R_{\rm BLR}$, the energy density of the
line photons as seen in the comoving frame becomes (up to a factor of order unity):
\begin{equation}
U^\prime_{\rm BLR} \, \sim \, \Gamma^2 {L_{\rm BLR} \over 4\pi R^2_{\rm BLR}c} \, =
\, {\Gamma^2 \over 12\pi}
\end{equation}
where we have used $R_{\rm BLR} \sim 10^{17} L^{1/2}_{\rm d, 45}$ cm, and $C=0.1$.
A similar relation holds for the IR radiation reprocessed by the torus located at a distance
$R_{\rm IR}$, namely when $R_{\rm BLR}<R_{\rm diss}<R_{\rm IR}$.
This limits $R_{\rm diss}$.

\vskip 0.3 cm
\noindent
{\it Magnetic field} --- 
The magnetic energy density, although is formally a free parameter, must satisfy
the Compton to synchrotron luminosity ratio, i.e.
$L_{\rm C} /L_{\rm syn} = U^\prime_{\rm rad} /U^\prime_B$. $U^\prime_{\rm rad}$ includes
both internally produced radiation (i.e. by synchrotron) and radiation produced
externally (directly by the disc or reprocessed and re--isotropized by the BLR and the torus).

\vskip 0.3 cm
\noindent
{\it Bulk Lorentz factor} --- 
The value of $\Gamma$ determines the value of the radiation energy
density of the external seed photons ($\propto \Gamma^2$) and hence
the value of the magnetic field required to have the observed
synchrotron to inverse Compton luminosity ratio.
Further information come  from the peak frequencies of the synchrotron and
inverse Compton components, that depend also on the break energy of the
electron distribution.

\vskip 0.3 cm
\noindent
{\it Injected power} --- The power is injected throughout the source
in the form of relativistic electrons. Through the continuity equation
we calculate the particle distribution as a result of injection,
cooling and possible pair production.  The total injected power is
such that the radiation produced by these particles agrees with the
observed data.  The injected distribution is assumed to be a smoothly
broken power law (see Eq. \ref{qgamma} in the Appendix). 
The resulting distribution, modified by cooling,
must agree with the observed slopes.

%--------------------------------------------------
\begin{figure} 
\vskip -0.5 cm
\hskip -0.3cm
\psfig{file=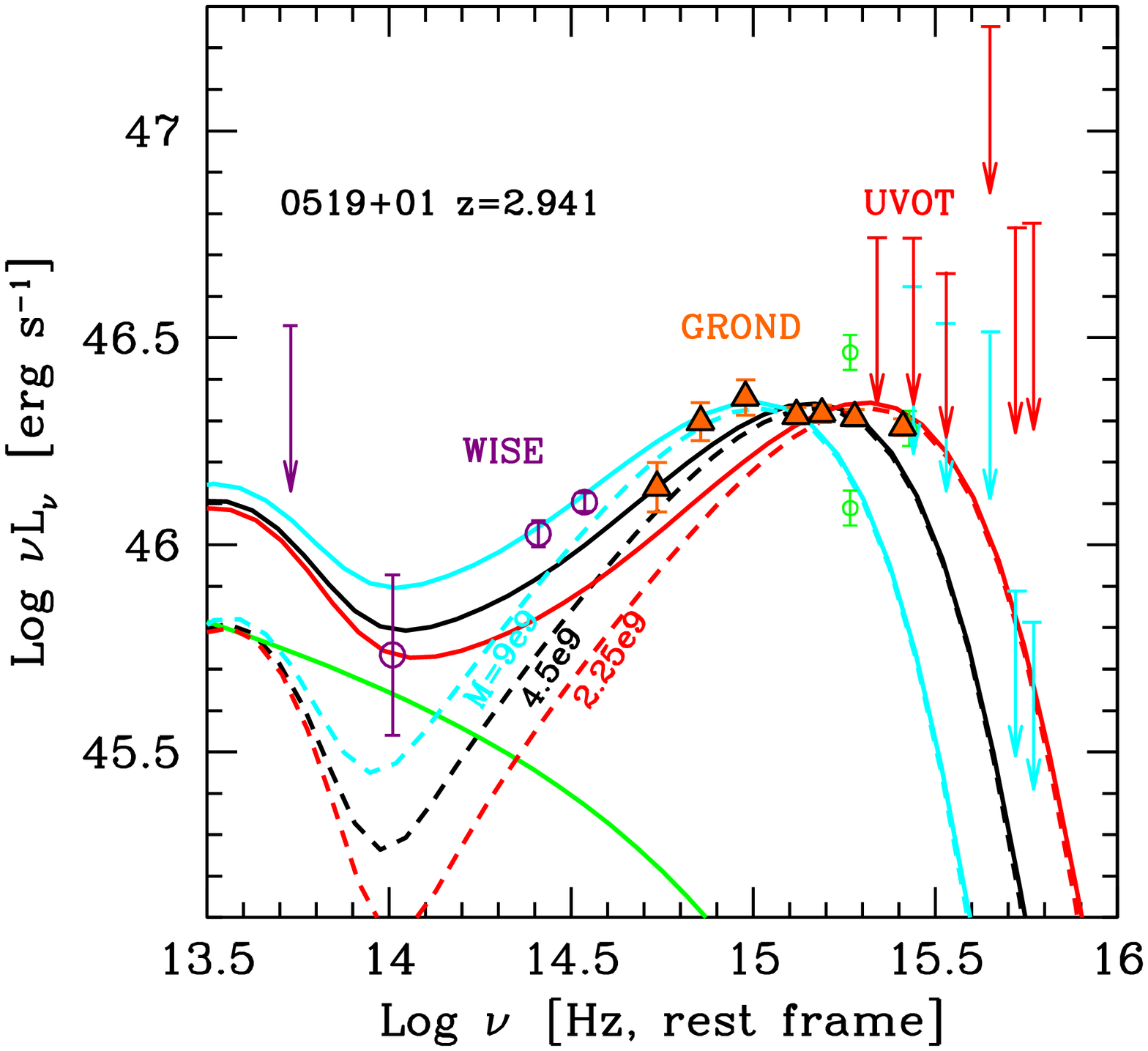,height=8.5cm,width=9cm}
\vskip -1 cm
\hskip -0.3cm
\psfig{file=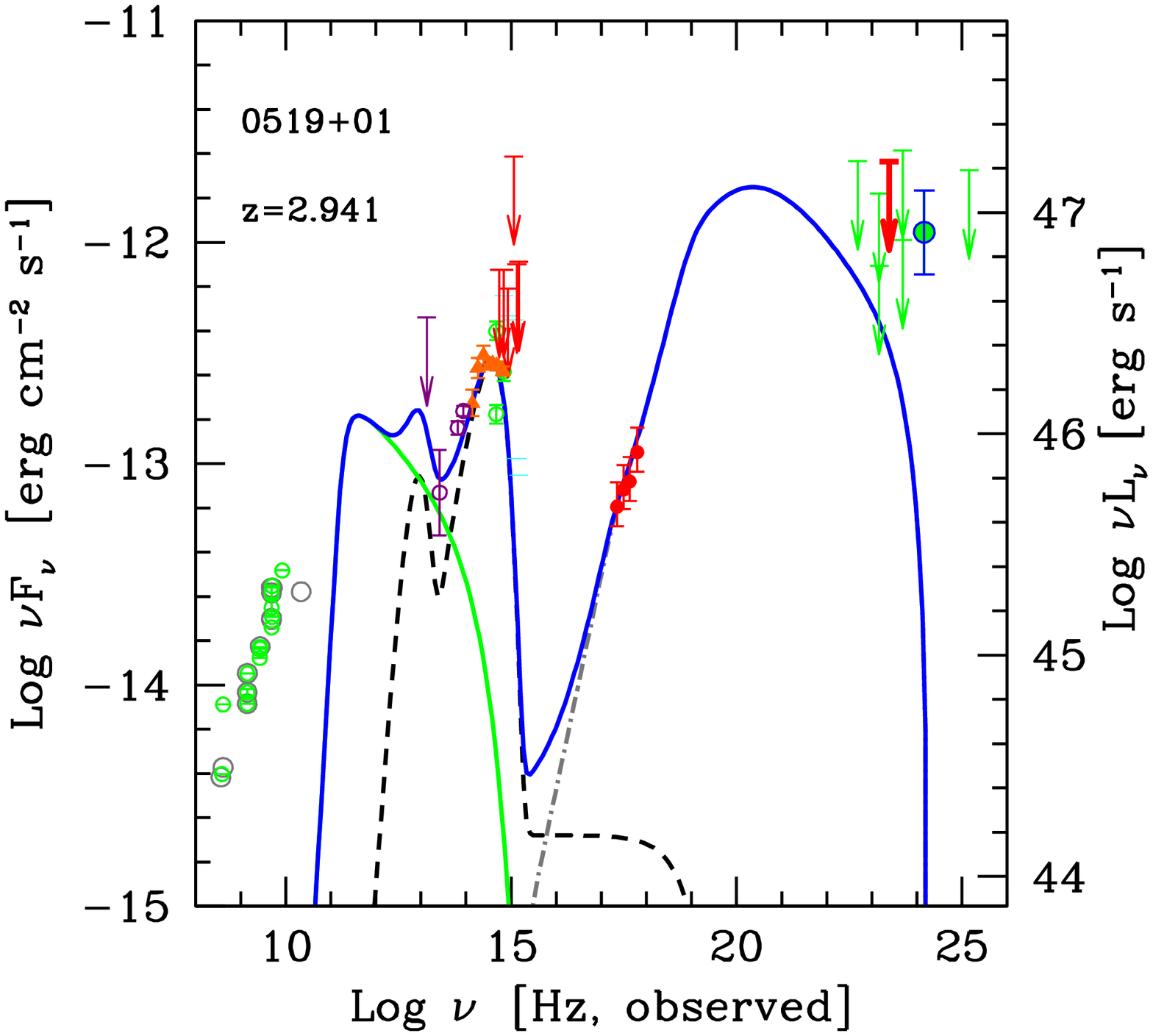,height=8.5cm,width=9cm}
\vskip -0.5 cm
\caption{ 
Top panel: a zoom of the SED of the blazar PKS 0519+01
in the IR, optical and UV. Light (cyan) arrows are the upper limit
as observed by UVOT, darker (red) arrows are the same data
de--absorbed by the predicted amount of Ly$\alpha$ forest absorption 
along the line of sight (see Ghisellini et al. 2010a).
Triangles (orange) are the GROND data, empty circles are the WISE data.
We also show three accretion disc models  (short dashed lines) with
the same accretion luminosity and three different black hole mass, 
from 2.25 to 9 billion solar masses (as labelled), together with the
contribution of the torus emission, emitting in the IR.
The thin solid (green) line is the synchrotron component, and
the solid (cyan, black and red) lines are the sum of the accretion 
disc+torus+synchrotron
flux. 
Note that {\it WISE} data are not simultaneous.
Bottom:
The entire SED of the PKS 0519+01, together with the
adopted model.
Filled (red) circles are the XRT data, and the (heavy) arrow
in the $\gamma$--ray band is the upper limit 
corresponding to 1 month of {\it Fermi} data 
centered to the {\it Swift}+GROND observations,
while the other arrows in the $\gamma$--ray band correspond
to data from 2yr integration. 
Note also the detection at $\nu\sim 10^{24}$ Hz.
Archival data are form the online service of ASI Science Data Center
(ASDC, green filled circles) and NED (empty circles).
Short dashed line: contribution from the accretion disc (with 
a black hole mass $M=4.5\times 10^9 M_\odot$), IR torus and corona.
Thin solid (green) line: synchrotron; dot--dashed line: EC component.
Solid (blue) line: sum of all components.
} 
\label{0519}
\end{figure}
%--------------------------------------------------

\subsection{Caveats}

Within the framework of the adopted model, there is some degeneracy
between a few set of parameters, that can be broken if some additional
information, besides the SED, is provided.
For instance, the bulk Lorentz factor and the viewing angle,
together with the injected power in relativistic particles, can
have a range of possible values. 
If, in addition, we have a limit on the variability timescale and/or
the superluminal speed, then we can chose a unique set of parameters.

The black hole mass found by fitting a Shakura--Sunjaev 
disc to the near IR--optical--UV data gives excellent results if the flux
at these frequencies is not contaminated by the synchrotron flux.
Otherwise, the disc luminosity can be estimated rather accurately from
the observed broad line luminosities, but the black hole mass
can have a large uncertainty, partly mitigated by assuming that the
disc cannot be super--Eddington (thus yielding a lower limit to the 
black hole mass).

Also the derived jet powers bear some uncertainties due to several
unknowns:
i) we do not know if charge neutrality is provided by protons, or by
positrons. Recent studies (Ghisellini \& Tavecchio 2010) have shown that
a pure pair plasma would suffer a severe Compton drag while crossing
the broad line region, limiting the positron/proton ratio to nearly 20
(in agreement with independent estimates put forward by  Sikora \& Madejski 2000);
ii) we can estimate the amount of {\it emitting} particles, but there can
be additional particles that are not accelerated, but nevertheless participate
to the bulk motion of the jet, hence to its kinetic power;
iii) the estimated magnetic field $B$ is the one in the emitting region. If
the dissipation mechanism is magnetic reconnection, it is likely that 
in the emitting region $B$ is smaller than in the surroundings.

Bearing in mind these limitations, we now discuss the physical parameters
found by adopting our model.

\subsection{Physical parameters}

Tab. \ref{para} lists the parameters of the applied model.
It also reports the parameters adopted in Ghisellini et al. (2011)
for the two sources in common with that paper.
Consider that the SED available at the time of Ghisellini et al. (2011)
was largely incomplete, lacking the {\it WISE}, GROND and {\it Swift} data.
Moreover, also the high frequency radio coverage has been improved with data provided
by the {\it Planck} and {\it WMAP} satellites (see also Giommi et al. 2012 for
a collection of blazar's SED including {\it Planck} and {\it WMAP} data).
The improved characterization of the SED of these blazars allowed
a better estimate of the physical parameters: we find that
although the values found now are not very different from what we have
guessed before, the uncertainty is much less.

In general, all the derived parameters are well within the distributions
derived for a large sample of $\gamma$--ray loud blazars studied in 
Ghisellini et al. (2010a).
For all 5 blazars in our sample,
the region dissipating most of the flux we see is located at several hundreds
of \sch\ radii from the black hole, with bulk Lorentz factors in the range 10--15, 
and small viewing angle ($\theta_{\rm v}\sim 3^\circ$). 
The magnetic field is in the range 1--8 Gauss, and the intrinsic power injected in the form 
of relativistic electrons is of the order of $10^{44}$ erg s$^{-1}$, as measured
in the comoving frame.
The black hole mass is around $M\sim 10^9 M_\odot$, with a luminosity
of the accretion disc ranging from $10^{45}$ erg s$^{-1}$ (for PKS 1348+007)
to $\sim$ (2--3)$\times 10^{46}$ erg s$^{-1}$ (for PKS 0519+01 and TXS 1149--084).
These are all very typical values for FSRQs.
In the following we discuss in more detail individual sources.

\subsection{PKS 0519+01}   % Lblr=44.31 from Lgamma

Present in the 2LAC catalog, this $\gamma$--ray blazars had no 
previous X--ray information.
There is no redshift in NED, the value for $z$ comes from the value listed in the 2LAC catalog,
but the spectrum is still unpublished.
% In the optical, there are a few data from the United States Naval Observatory (USNO).
Our observations (and the {\it WISE} data) greatly improve
our knowledge of the SED, as shown in Fig. \ref{0519}, despite the fact 
that for this source, the UVOT data were only upper limits.
Note that for each UVOT upper limit we plot two arrows,
corresponding to the observed datum and the de--absorbed one.
The latter is derived assuming, along the line of sight, an average distribution 
of absorbing Ly$\alpha$ clouds (see Fig. 3 in Ghisellini et al. 2010a).

The GROND photometric points allow to determine the peak
of the accretion disc spectrum, hence its luminosity 
$L_{\rm d}\sim 2\times 10^{46}$ erg s$^{-1}$ and  black hole mass,
that turns out to be $M=4.5\times 10^9 M\odot$. 
This is the largest value we find for the 5 blazars here considered.
To estimate the uncertainty for this value, we show in Fig. \ref{0519}
the fit with a black hole mass of 2.25, 4.5 and 9 billion solar masses,
as labelled, keeping the same $L_{\rm d}$.
The fit with the largest mass underestimates the high frequency GROND point, 
while the fit with the lowest mass underestimates all but the high frequency
GROND fluxes.
We can conclude that the mass determination has an uncertainty, in this case,
of less than a factor 2.

The bottom panel of Fig. \ref{0519} shows the complete SED of the source together
with the model.
We show, separately, the synchrotron component (solid light green line)
the torus+disc+X--ray corona contribution (black short--dashed line) and
the inverse Compton contribution, dominated by scattering with 
emission line seed photons (grey dot--dashed line).
The thick (blue) solid line is the sum.
We show also the {\it Fermi} upper limit on the $\gamma$--ray flux resulting from
a month of data centered on the time of the {\it Swift} and GROND observations
(thick red arrow).

%--------------------------------------------------
\begin{figure} 
\vskip -0.5 cm
\hskip -0.3cm
\psfig{file=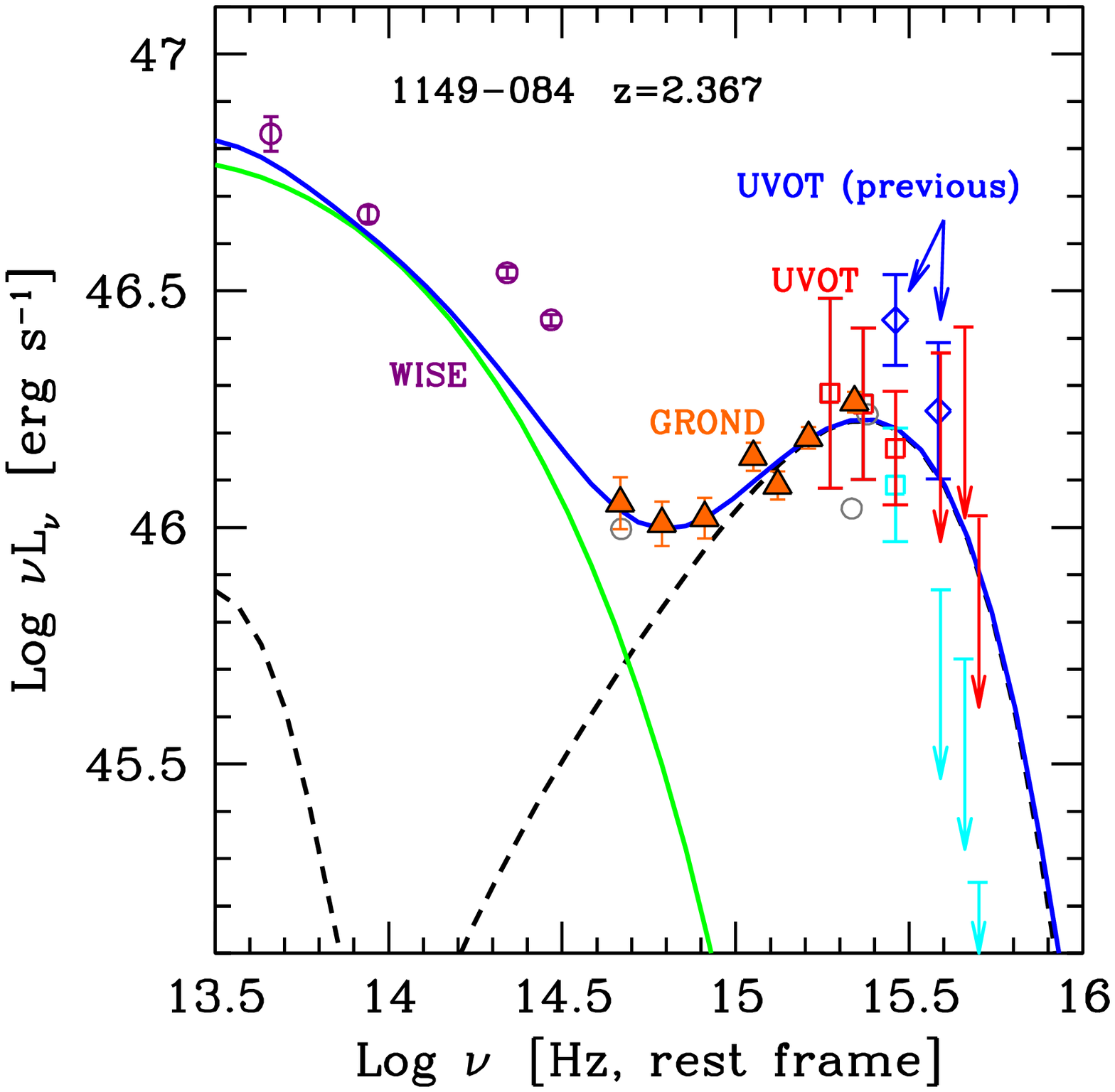,height=8.5cm,width=9cm}
\vskip -1 cm
\hskip -0.3cm
\psfig{file=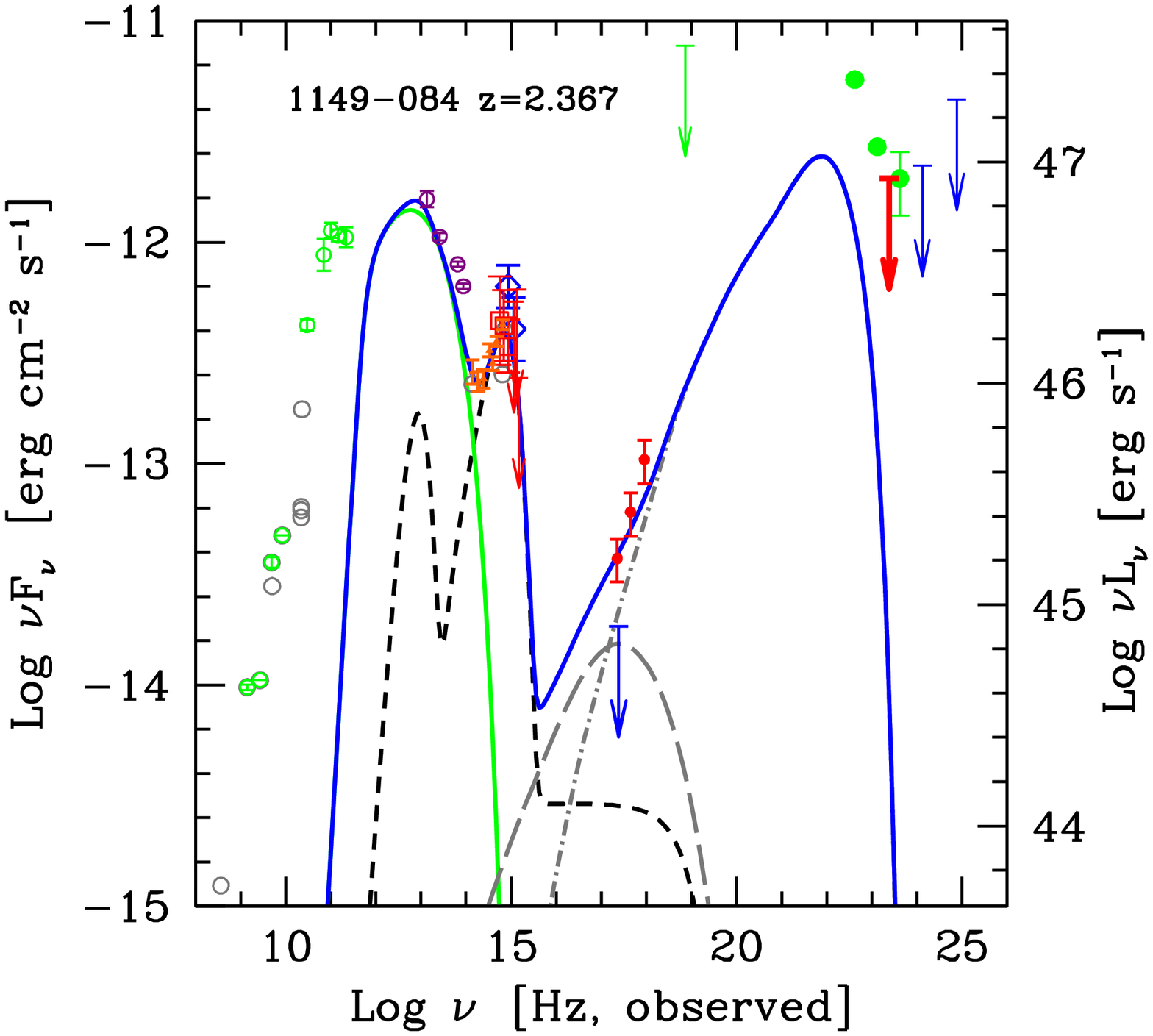,height=8.5cm,width=9cm}
\vskip -0.5 cm
\caption{Lines and symbols as in Fig. \ref{0519}, for TXS 1149--084.
Bottom panel: in this case the SSC component (long dashed grey line)
contributes to the soft X--ray flux.
The upper limit in X--rays (blue arrow) corresponds
to an earlier and very short (1 ks) {\it Swift} observation,
when the source was detected in two UVOT filters 
as labelled in the top panel [``UVOT (previous)"].
} 
\label{1149}
\end{figure}
%--------------------------------------------------

\subsection{TXS 1149--084}     % Lblr=44.96 from Lgamma
% da Shaw M.S., Romani R.W., Cotter G. et al. 2012 ApJ 748, 49:
% AR    dec   z       F5980 err  al   err  L3000  err   LMgII  err   FWHM  err  M    err  L1350  err   LCIV    err   FWHM    err M      err  Telescope
% J1152 -0841 2.367     4.3 0.0 -0.07 0.35 0      0     0      0      0     0   0    0    46.130 0.020 44.301  0.327 7200   2800 9.38   0.77 HET    
This source has been observed spectroscopically in the optical by
Shaw et al. (2012), that reported a 
luminosity ($L_{\rm CIV}=2\times10^{44}$ erg s$^{-1}$) and a FWHM (7200 km s$^{-1}$)
of the CIV broad emission line, together 
with the luminosity of the continuum at 1350 \AA\ ($L_{1350}=1.3\times 10^{46}$erg s$^{-1}$).
These data allowed Shaw et al. (2012) to estimate a black hole mass
of $M=2.4\times 10^9 M_\odot$, applying the virial method.
Using the template of Francis et al. (1991) and Vanden Berk (2001), 
one can derive the overall luminosity of the BLR, $L_{\rm BLR} \sim 2\times 10^{45}$ 
erg s$^{-1}$ and then a disc luminosity ten times greater 
(assuming a covering factor equal to 0.1).
This value agrees well with the GROND+{\it Swift} data,
from which we determined the peak of the disc component,
with $L_{\rm d}=3.2\times 10^{46}$ erg s$^{-1}$.
Therefore also in this case the black hole mass is well determined,
$M= 1.5\times 10^9 M_\odot$.

In this source the {\it WISE} data, together with the high frequency
radio data (from {\it Planck}) show a strong synchrotron component,
peaking in the submm range.
This is also indicated by the GROND data, showing un upturn towards the low frequencies.
This upturn constrains the possible models capable of reproducing the synchrotron peak.
The self--absorption frequency of our
compact emitting zone occurs at $\sim$ 760 GHz (observed frame),
making the synchrotron component very narrow.
We derive a rather large magnetic field ($\sim$8 G), to account
for the strength of the synchrotron flux. 
Accordingly, also the synchrotron Self Compton flux is not negligible,
and contributes to the soft X--rays (long dashed grey curve in Fig. \ref{1149}).

Also for this source we show the {\it Fermi}/LAT upper limit measured
from 1--month of data around the time of the {\it Swift} and GROND observations.

Tab. \ref{para} lists also the parameters used in Ghisellini et al. (2011),
for which no GROND data were available and there was only an upper limit to the X--ray flux
(shown as a blue arrow in Fig. \ref{1149}), implying that the source
has brightened in the X--ray band.
The main differences with those results concern the black hole mass 
(it was $\sim$3 times greater), the value of $R_{\rm diss}$ (4 times larger) 
and the magnetic field (5.5 times smaller).
The previous UVOT fluxes were slightly larger, and in the absence of
additional optical IR data these resulted in an overestimation of the
black hole mass and and disc luminosity (see Tab. \ref{para}),
instead of a larger synchrotron flux.
This well illustrates the importance of having a good
coverage in the IR--optical, and also some information on the 
emission lines (from Shaw et al. 2012).
The better coverage in the near and far IR and in the submm
range allows to characterize better the synchrotron component, 
while the detection in the X--rays (at a flux larger than the previous
upper limit) allows to determine the importance of the inverse Compton
process (sum of SSC and EC).
We find that the overall SED can be explained
assuming a relatively strong synchrotron (and SSC) components,
consequence of a magnetic field larger than the one assumed in
Ghisellini et al. (2011).

%---------------------------------
\begin{table} 
\centering
\begin{tabular}{llllll}
\hline
\hline
Name   &$\log P_{\rm r}$ &$\log P_{\rm B}$ &$\log P_{\rm e}$ &$\log P_{\rm p}$ &$\log P_{\rm j, min}$ \\
\hline   
0519+01    &45.00 &45.68 &44.53 &46.94 &45.30\\  
1149--084  &45.08 &46.22 &43.93 &46.16 &45.38\\
~          &{\it 45.47} &{\it 45.87} &{\it 43.83} &{\it 46.24} &{\it 45.77} \\
1333+2725  &45.18 &44.58 &44.33 &45.96 &45.48\\
1344--1723 &45.54 &45.23 &44.07 &46.12 &45.84 \\
~          &{\it 45.66} &{\it 44.74} &{\it 44.43} &{\it 46.03} &{\it 45.96} \\
1348 high  &45.00 &44.52 &43.96 &45.38 &45.30 \\
1348 low   &44.68 &44.12 &43.59 &45.34 &44.98\\
\hline
\hline 
\end{tabular}
\vskip 0.4 true cm
\caption{
Logarithm of the jet power in the form of radiation ($P_{\rm r}$), 
Poynting flux ($P_{\rm B}$),
bulk motion of electrons ($P_{\rm e}$) and protons ($P_{\rm p}$,
assuming one proton per emitting electron). 
The last column lists the minimum jet power, calculated assuming
that the radiation drag of the jet halves its bulk Lorentz factor.
This limit corresponds to twice the radiated power $P_{\rm r}$.
Powers are in erg s$^{-1}$.
The parameters in italics refer to the physical quantities found in Ghisellini et al. (2011).
}
\label{powers}
\end{table}
%---------------------------------

%--------------------------------------------------
\begin{figure}
\vskip -0.5 cm
\hskip -0.3cm
\psfig{file=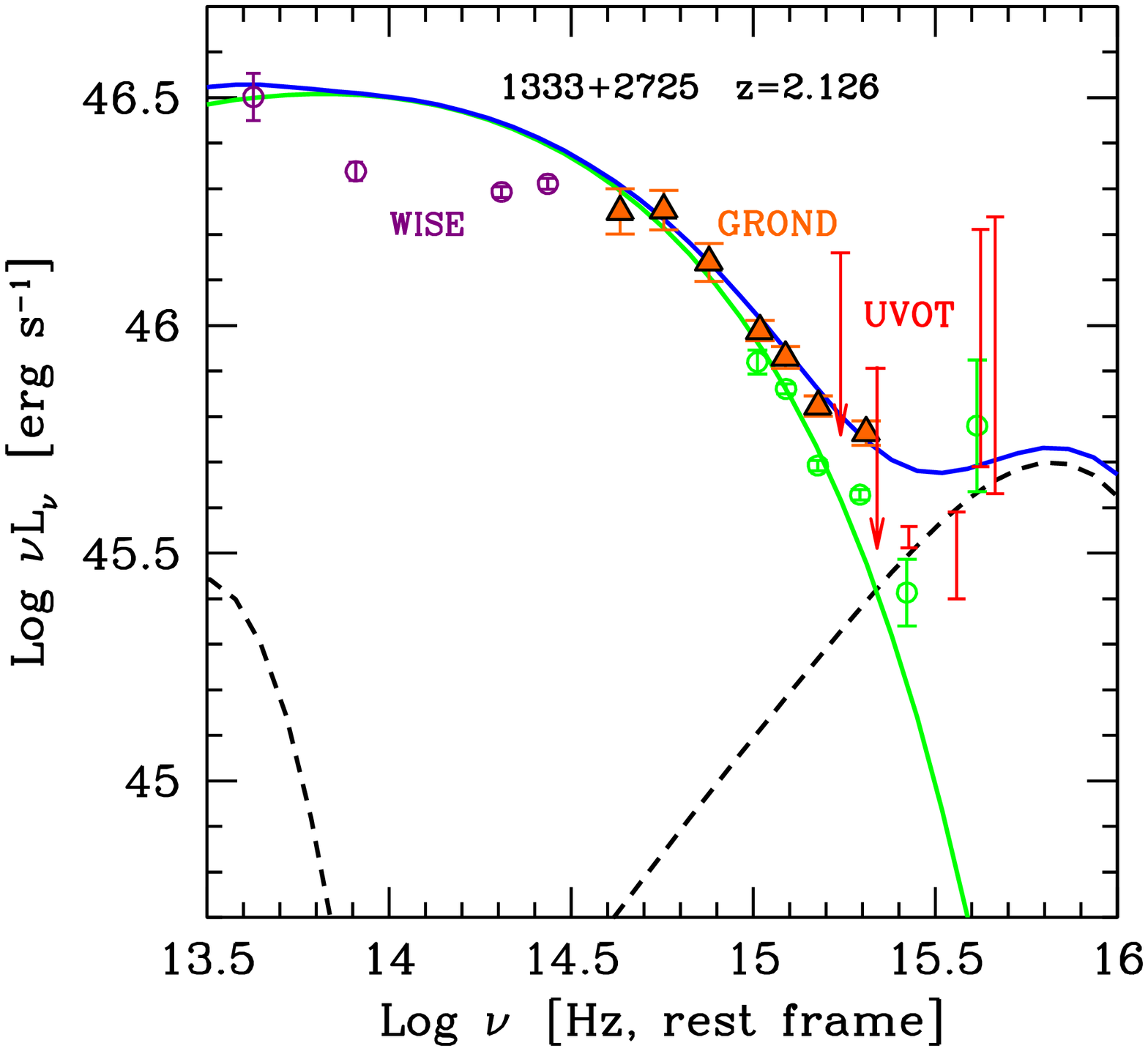,height=8.5cm,width=9cm}
\vskip -1 cm
\hskip -0.3cm
\psfig{file=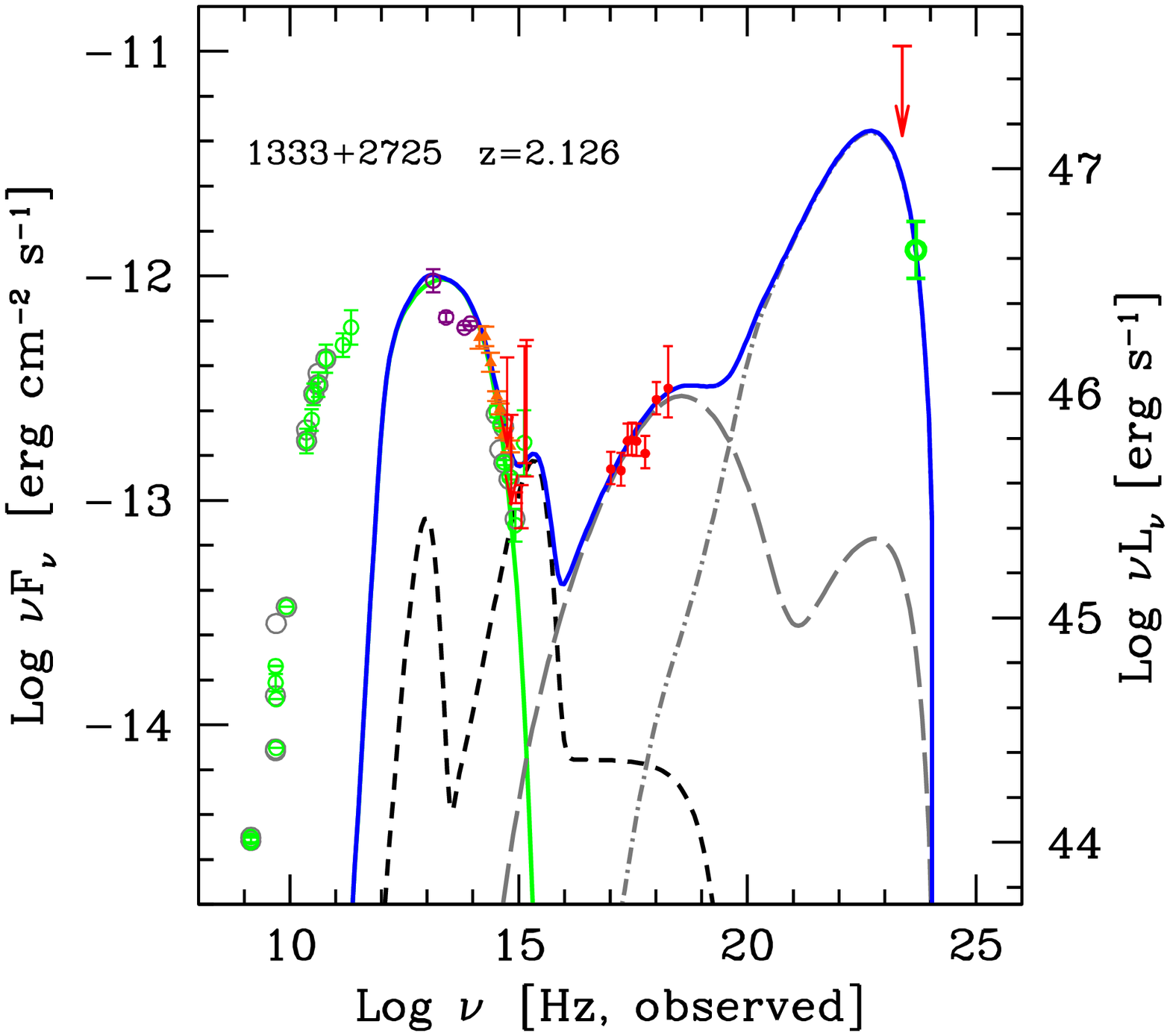,height=8.5cm,width=9cm}
\vskip -0.5 cm
\caption{Lines and symbols as in Fig. \ref{0519}, for MG2 J133305+2725. 
Also in this case, as for TXS 1149--084, the SSC component
contributes to the soft X--ray flux.
The bars in the bluest UVOT filters indicate
the possible range of intrinsic flux levels: the lowest 
extremes correspond to the detected flux, and the highest extremes
to the flux once de--absorbed by the average intervening Ly$\alpha$
absorption as calculated in Ghisellini et al. (2010).
} 
\label{1333}
\end{figure}
%--------------------------------------------------

\subsection{MG2 J133305+2725}    % Lblr=44.59 from Lgamma

This is a blazar that is present in the photometric optical SDSS survey (with a
magnitude $r=20.18$), but with no spectroscopic observations.
UVOT detected the source in the bluest filters.
At these frequencies the flux is partially absorbed by intervening
Ly$\alpha$ clouds, whose total optical depth can be roughly calculated
by averaging over many line of sights (as done in Ghisellini et al. 2010).
There is however a large dispersion around these mean values,
and for this reason we have indicated, with bars, the possible range
of the de--absorbed UVOT fluxes.

The slope defined by the GROND data is steep, not consistent 
with a disc spectrum, that must therefore be hidden by the tail of the
synchrotron flux.
A strong synchrotron component is indeed needed by the {\it WISE}, {\it WMAP} and
{\it Planck} data.
However, for the fit, we have given less weight to the {\it WISE} data, since
they are not simultaneous.
% Despite the high redshift of the source,
% UVOT detected it in the bluest filters, where we expect the
% absorption of the intervening Ly$\alpha$ clouds.
% Since the estimate of the latter is affected by a large
% uncertainty, for this blazar we have chosen to show the observed 
% and the de--absorbed values joint by a bar.

The emission disc component shown in Fig. \ref{1333} is only illustrative, 
but its luminosity cannot much be less than shown, due to the presence
of the broad emission lines that make this blazar a FSRQs (but there is no published 
information on the line strength).
For the black hole mass we have no strong constraints but
a slight flattening of the GROND data towards the blue,
possibly indicating an upturn of the spectrum.
If due to the presence of the accretion disc, this implies a relatively small mass
(i.e. a high maximum temperature), so we have chosen
an illustrative value of $M=10^8 M_\odot$. 
For this mass the disc must emit at 60\% the Eddington rate.

A strong synchrotron component implies that the SSC flux can contribute to the
soft X--ray spectrum, as shown in Fig. \ref{1333}, while the EC component 
dominates the bolometric output as in the other sources.

% da Shaw M.S., Romani R.W., Cotter G. et al. 2012 ApJ 748, 49:
% AR    dec   z       F5980 err  al   err  L3000  err   LMgII  err   FWHM  err  M    err  L1350  err   LCIV    err   FWHM    err M      err  Telescope
% J1344 -1723 2.506     2.7 0.0 -2.20 0.22 0      0     0      0      0     0   0    0    45.770 0.046 44.076  1.592 6000  11200 9.12   3.78 P200   :
% Lblr= L_CIV * 555.8/63 --> log Lblr=0.9456 + log L_CIV
% Lbol=5.15*L3000      Lbol=3.81*L1350 
% 1152=1149: Log Lblr= 45.247 Lbol=46.71;   1344: Log Lblr= 45.022; Lbol=46.35
% model: 1149: Ldisc = 2.5e46 (log=46.4), M=1.5e9 (log=9.18)  1344: Ldisc=1.4e46 (Log=46.12), M=1.9e9 (log=9.28)

%--------------------------------------------------
\begin{figure}
\vskip -0.5 cm
\hskip -0.3cm
\psfig{file=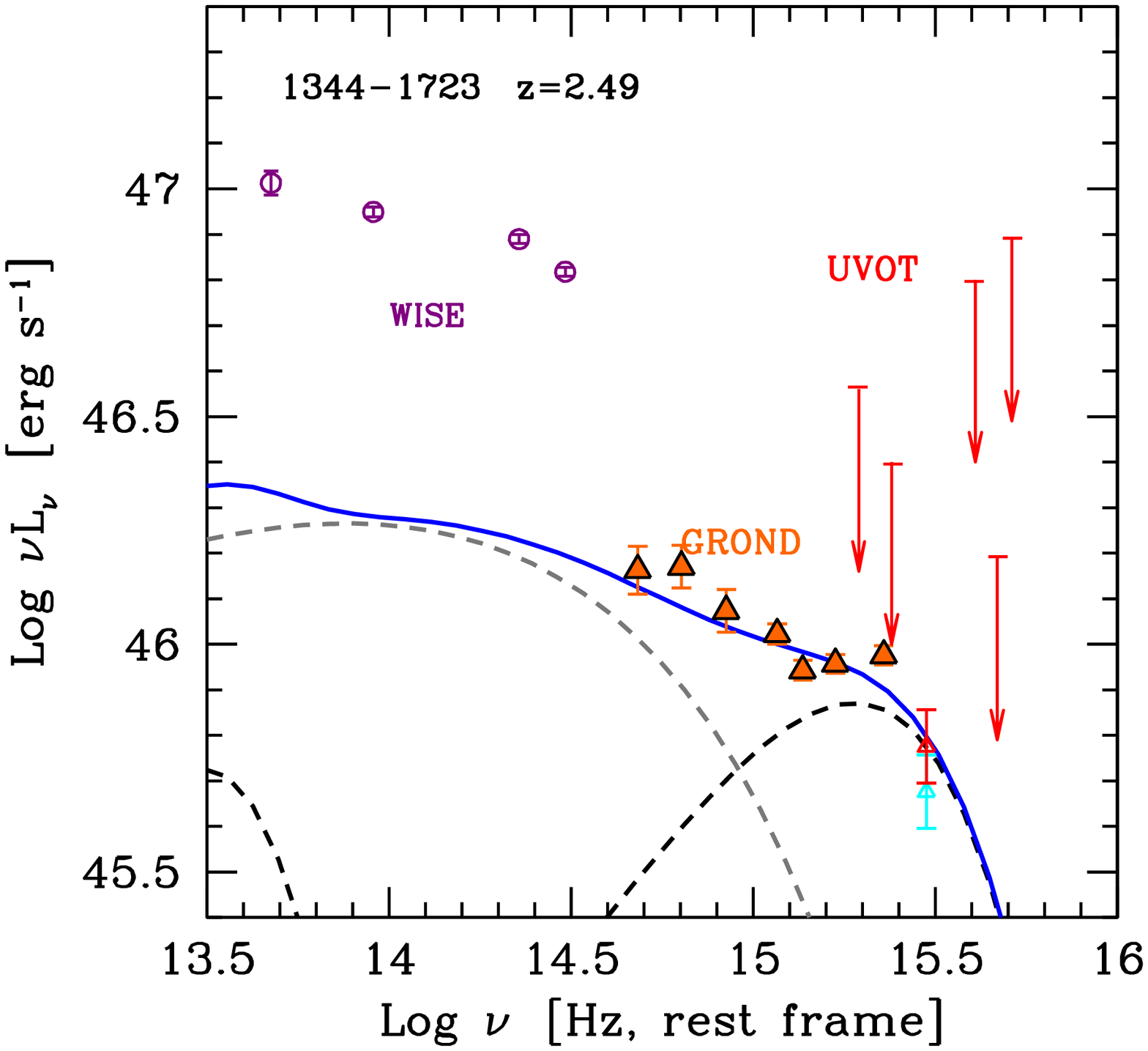,height=8.5cm,width=9cm}
\vskip -1 cm
\hskip -0.3cm
\psfig{file=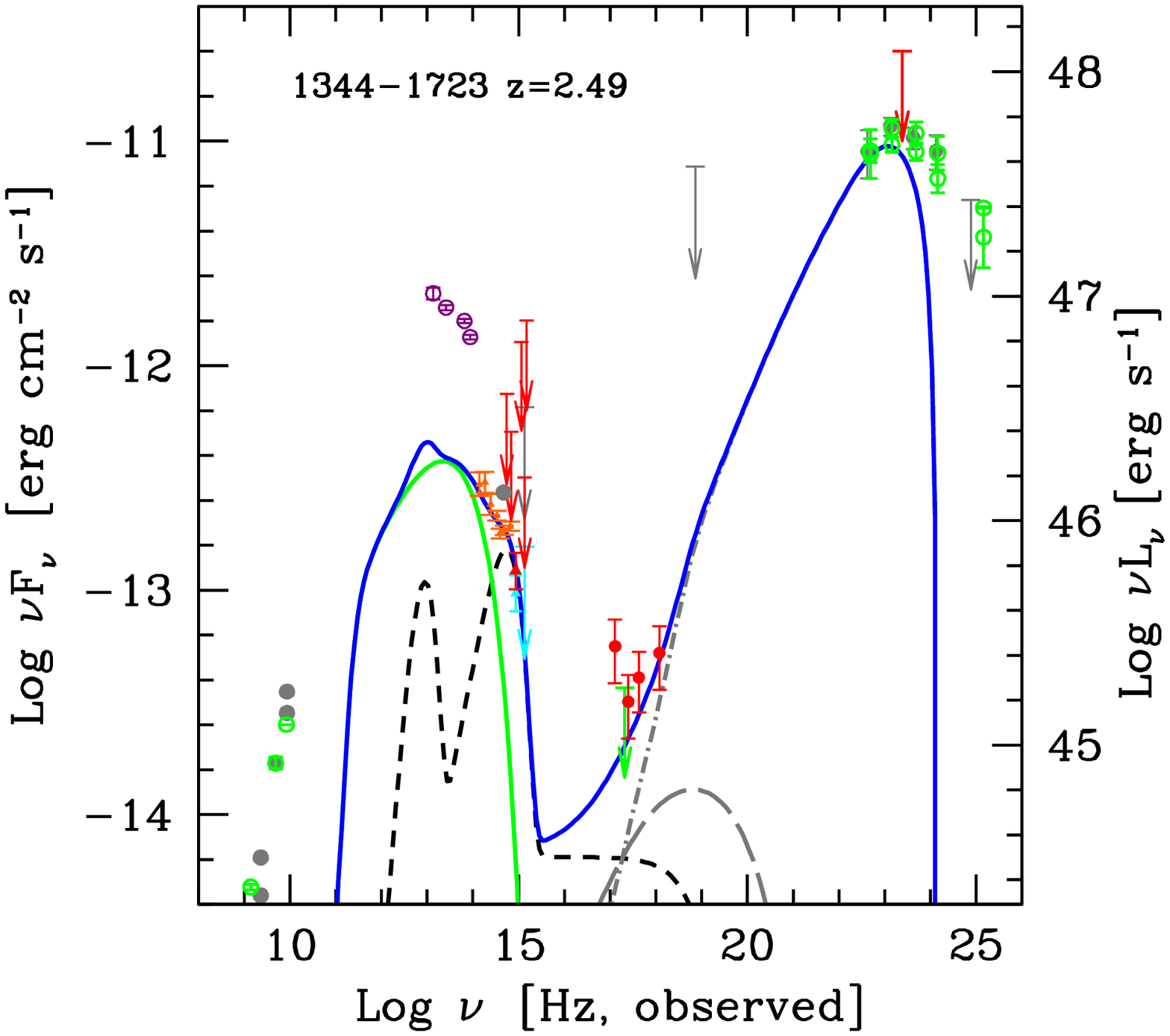,height=8.5cm,width=9cm}
\vskip -0.5 cm
\caption{Lines and symbols as in Fig. \ref{0519}, for PMN J1344--1723.
} 
\label{1344}
\end{figure}
%--------------------------------------------------

%--------------------------------------------------
\begin{figure}
\vskip -0.5 cm
\hskip -0.3cm
\psfig{file=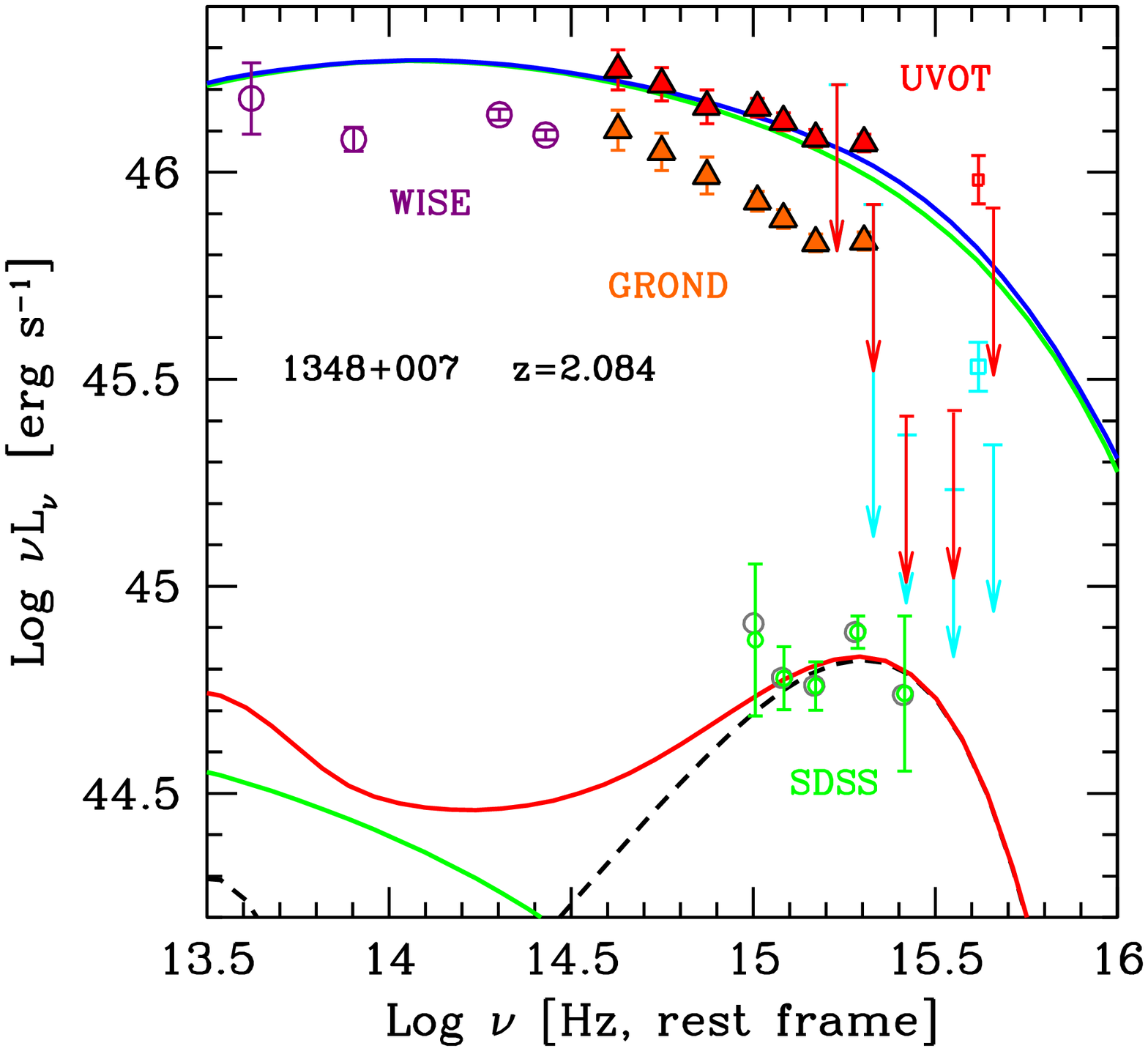,height=8.5cm,width=9cm}
\vskip -1 cm
\hskip -0.3cm
\psfig{file=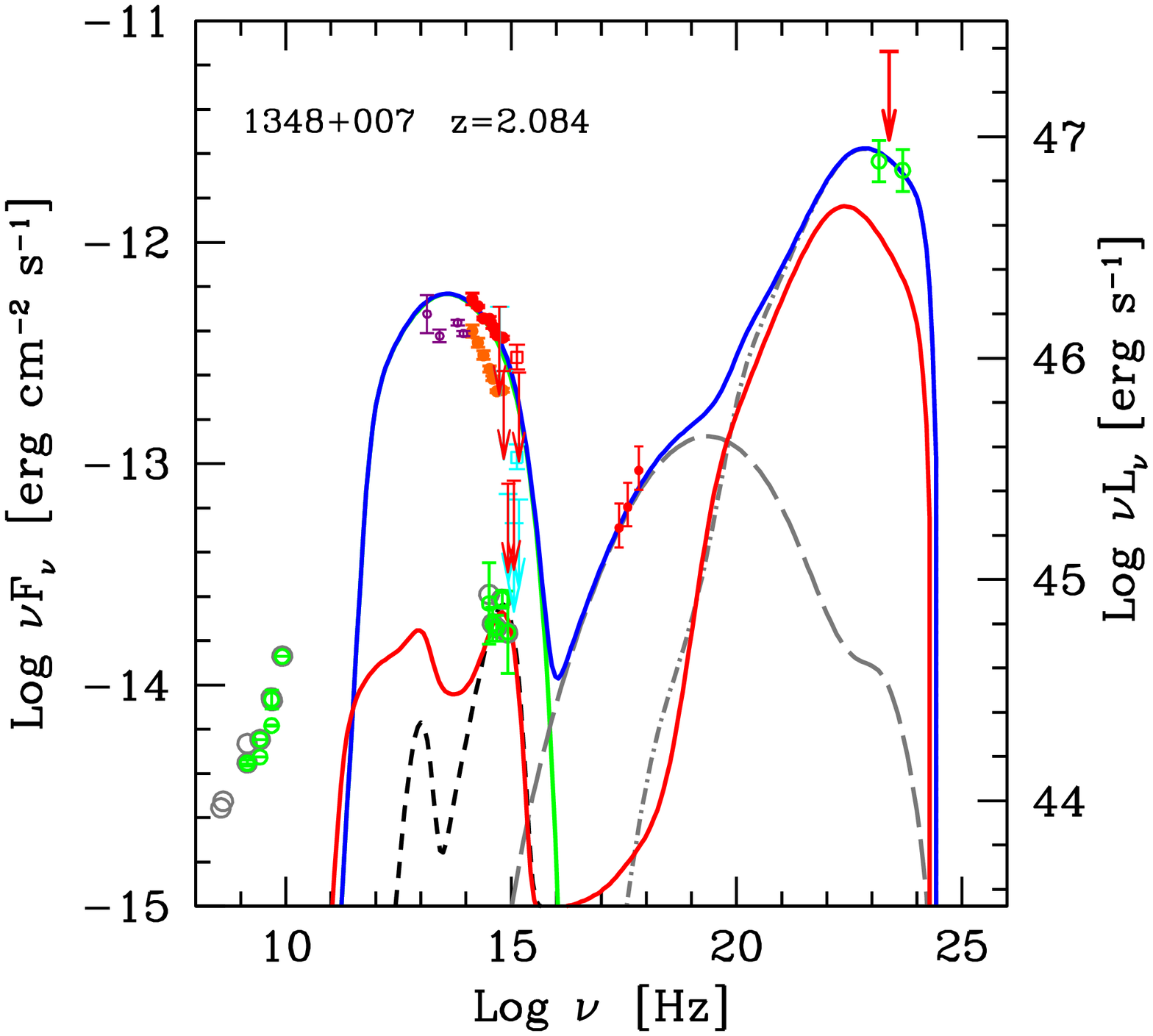,height=8.5cm,width=9cm}
\vskip -0.5 cm
\caption{
Lines and symbols as in Fig. \ref{0519}, for PKS 1348+007.
Note the large amplitude variability in the optical.
The SDSS photometric fluxes, taken on May 20, 2009,
are a factor $\sim$30 below the GROND+{\it Swift}+WISE data.
Fast variability is present also in our GROND data,
taken two nights apart (see Tab. \ref{grondlog}, Tab. \ref{grond}). 
} 
\label{1348}
\end{figure}
%--------------------------------------------------

\subsection{PMN J1344--1723}   % Lblr= 45.44 from Lgamma

This source has been observed spectroscopically in the optical by
Shaw et al. (2012), that derived the 
luminosity of the CIV broad line ($L_{\rm CIV}=10^{44}$ erg s$^{-1}$), corresponding
(using Francis et al. 1991 or Vander Berk et al. 2001), to $L_{\rm BLR}\sim 10^{45}$ erg s$^{-1}$.
Adopting a covering factor $C\sim 0.1$ we then have $L_{\rm d}\sim 10^{46}$ erg s$^{-1}$.
The GROND data show a flattening (in $\nu F\nu$) of the spectrum, that we
interpret as the emergence of the accretion disc component.
With UVOT we have upper limits in all filters except in the $u$ band.
Note that in $u$ the source is already affected by possible absorption
by intervening Ly$\alpha$ clouds. 
In Fig. \ref{1344} both the observed and the de--absorbed flux are plotted.
However, we caution about the large uncertainty connected with the use of an average
distribution of absorbers along the line of sight.
By decomposing the optical--UV emission with a synchrotron+accretion disc 
component, we derived $L_{\rm d}=1.2\times 10^{46}$ erg s$^{-1}$,
in agreement with what is indicated by the CIV broad line.
We then derive a black hole mass $M=1.5\times 10^9 M_\odot$.
This value is about the same as that derived by Shaw et al. (2012) using the virial
method, the FWHM of the CIV broad line (6000 km s$^{-1}$) and the continuum luminosity
at 1350 \AA\ ($L_{1350}=5.9\times 10^{45}$erg s$^{-1}$), giving 
$M=1.3\times 10^9 M_\odot$.

The {\it WISE} data indicate a strong synchrotron component.
However, these data cannot connect smoothly with the
GROND IR points, strongly suggesting that the synchrotron emission 
is variable with a large amplitude.
This implies also that the blazar was in a somewhat low state at
the epoch of our observations, and for this reason we did not attempt
to accurately reproduce the $\gamma$--ray flux.
For the model shown in Fig. \ref{1344} we have derived 2.3 Gauss for the 
magnetic field: with this value the SSC 
barely contributes to the soft X--rays.
At higher X--ray energies the flux is completely dominated
by the external Compton process (with line photons as seeds).
The present set of data can be compared with what was known
previously, and studied in Ghisellini et al. (2011, see their Fig. 5).
One can appreciate the great improvement, and consequently
the improved confidence on the derived physical quantities.

\subsection{PKS 1348+007}    % Lblr=44.87 from Lgamma

This blazar showed an extraordinary optical flare, as derived by comparing 
the optical SDSS data ($r=22.4$) with our GROND and UVOT data, together 
with the IR flux seen by {\it WISE}.
Fast variability is present also in our GROND data,
taken two nights apart (see Tab. \ref{grondlog}, Tab. \ref{grond} and
Fig. \ref{1348}). 
The source varied by about half a magnitude in all filters 
(except $K_s$), i.e. by $\sim$60\% in flux.
There is also a hint (although marginal) of a ``harder when brighter" behaviour.
Unfortunately, there are no SDSS spectra for this source.
Comparing our data with the SDSS photometric data, obtained on May 20, 2009,
the synchrotron emission had to vary by a factor $\sim$30.
As can be seen in Fig. \ref{1348}, the synchrotron component
completely outshines the disc emission at the time of our
GROND+{\it Swift} observations.
Fortunately, the SDSS data hint to the presence of an accretion disc, 
through a flat (in $\nu F_\nu$) slope suggested by the photometric data.
We have assumed that the SDSS fluxes are completely produced by
the accretion disc, and derived a luminosity $L_{\rm d}=1.3\times 10^{45}$
erg s$^{-1}$ and a black hole mass $M=4\times 10^8 M_\odot$.

The extraordinary optical variability of PKS 1348+007 is not
unprecedented, being similar to the optical flare shown
by 3C 454.3 in 2005 (Fuhrmann et al. 2006; Pian et al. 2006; 
Villata et al. 2006; Giommi et al., 2006).
In principle, these large optical variations could be
explained in two different scenarios.
In the first, one can assume that the jet power is unchanged,
but the dissipation regions varies  and is smaller
for the high optical state.
The magnetic field, which should decrease
along the jet, is larger for smaller $R_{\rm diss}$, 
implying a larger $U_B^\prime$.
On the other hand, the radiation energy density of the
broad line photons is constant within $R_{\rm BLR}$,
and in the comoving frame is $U^\prime_{\rm BLR} \sim \Gamma^2/(12\pi)$
(see Eq. 2).
Therefore the synchrotron to inverse Compton luminosity ratio $L_{\rm syn}/L_{\rm C}$
changes for varying $R_{\rm diss}$ even if the jet carries the same 
amount of power.
In this case the $\gamma$--ray luminosity could remain constant
or even decrease during an optical flare.
Alternatively, the jet power can vary, making the optical {\it and}
the $\gamma$--ray fluxes vary together.
In this case an high optical state should be accompanied by a
larger $\gamma$--ray flux.

Unfortunately, we do not have information about the high energy emission
for the low optical state of the source, so we cannot distinguish between
these two hypotheses.
We have simply explored the first option, looking for a solution for both states 
that maintains the total jet power roughly constant.
The two models shown in Fig. \ref{1348} correspond to the jet emission
produced at $R_{\rm diss}=60$ and 96 \sch\ radii, with $\Gamma=11$ and 15, respectively,
and with a equal power in bulk motion of the cold protons.
The magnetic field is 4.5 G for the high optical state (smaller $R_{\rm diss}$
and smaller $\Gamma$)
and 1.3 G. for the low optical state (larger $R_{\rm diss}$ and larger $\Gamma$).
The results demonstrate that this case is indeed possible.
Simultaneous observations in the X--rays and in the $\gamma$--ray range
when the source is in a low optical state can indeed decide if this is  what
really occurs.

%--------------------------------------------------
\begin{figure}
\vskip -0.5 cm
\hskip -0.3cm
\psfig{file=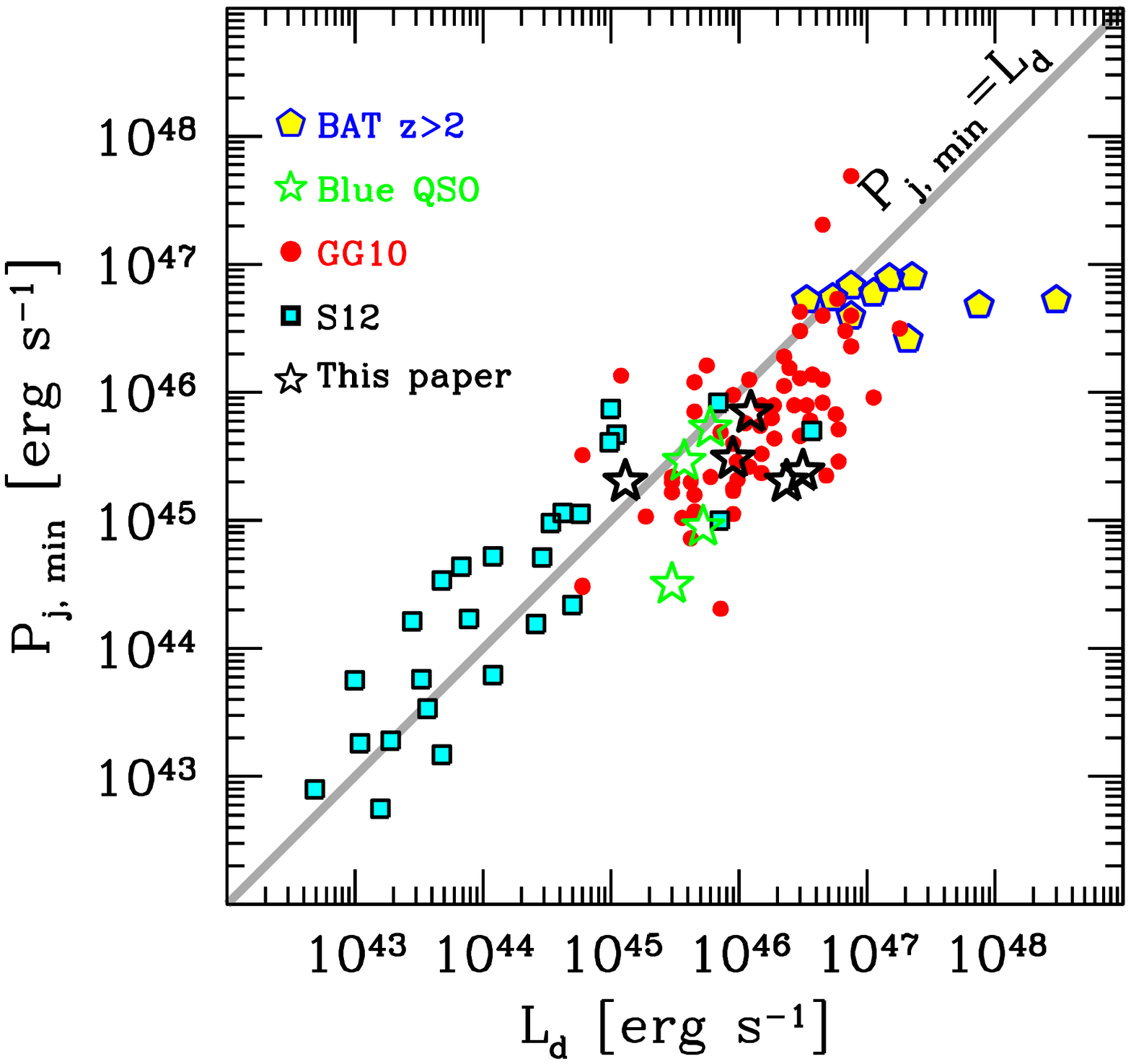,height=9cm,width=9cm}
\vskip -1 cm
\hskip -0.3cm
\psfig{file=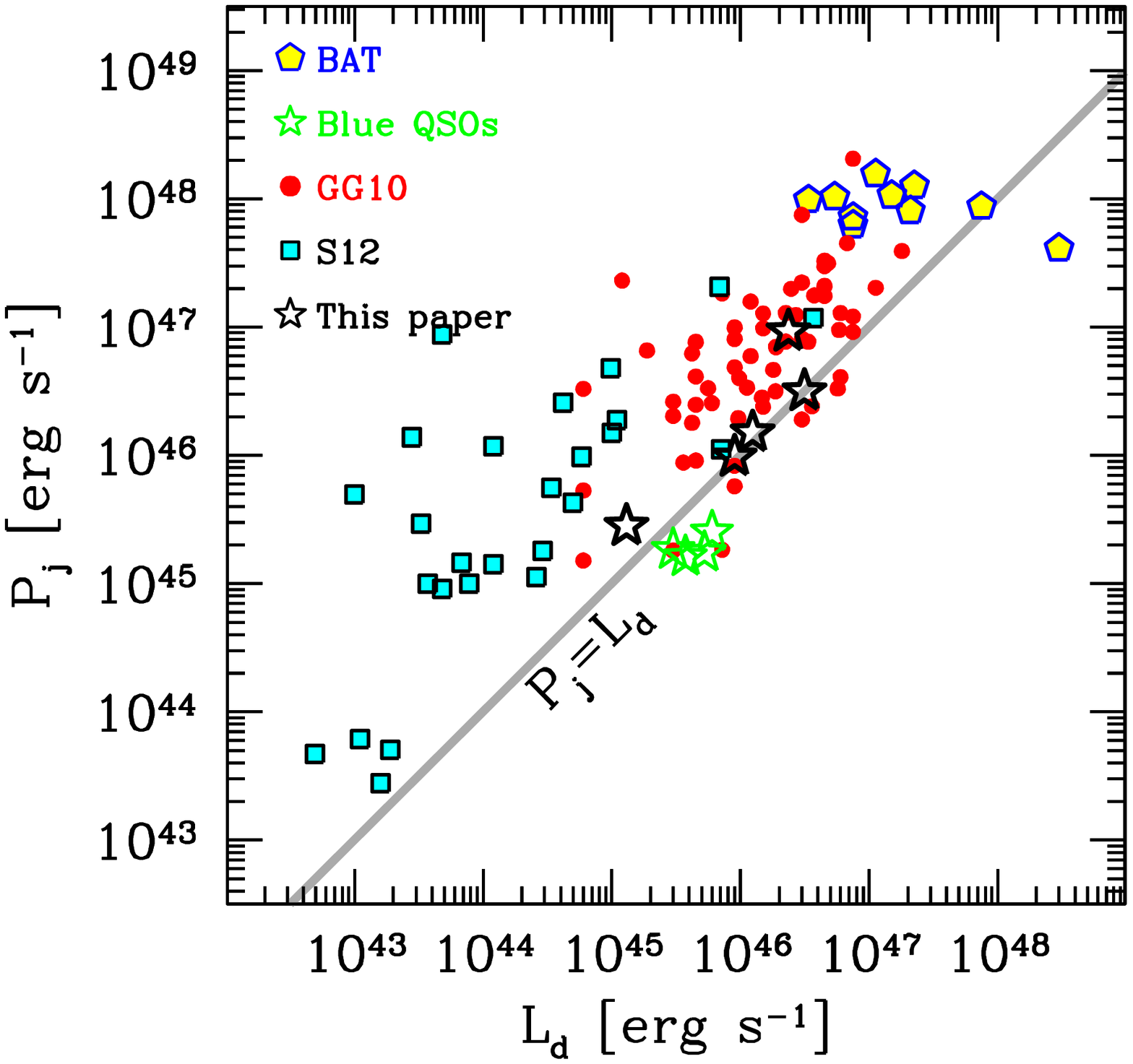,height=9cm,width=9cm}
\vskip -0.5 cm
\caption{The minimum jet power $P_{\rm j, min}$ (top panel)
and the total jet power $P_{\rm j}$ (including one proton per emitting electron),
as a function of the accretion luminosity $L_{\rm d}$.
Our sources (black stars) are compared with other $\gamma$--ray loud FSRQs
studied previously
(BAT $z>2$: Ghisellini et al. 2010a;
GG10: Ghisellini et al. 2010b;
Blue QSOs: Ghisellini et al. 2012;
S12: Sbarrato et al. 2012a).
The blazars studied in this paper lie on the bulk of the distribution.
The minimum jet power is twice the power that the jet
spends to produce the radiation we see (i.e. $P_{\rm j, min}=2P_{\rm r}$,
see text and Ghisellini \& Tavecchio 2010).
} 
\label{pjet}
\end{figure}
%--------------------------------------------------

\section{Discussion and conclusions}

With our {\it Swift}+GROND observational campaign, we
have secured the X--ray coverage for all the blazars
at $z>2$ present on the ``clean" 2LAC catalog.
The use of simultaneous GROND and {\it Swift} observations
were crucial to find the black hole mass and accretion rate
for 3 out of 5 sources (for MG2 J133305+2725
the synchrotron jet component was too strong
to see the accretion disc emission, and for 
PKS 1348+007 our observations cought the source
in a very high state, hiding the disc emission that
was instead visible by previous SDSS photometric observations).
We find that both $M$ and $L_{\rm d}/L_{\rm Edd}$ 
are not extreme, but rather standard for 
the powerful FSRQs detected by {\it Fermi}/LAT.
We have shown that the method of combining 
broad line luminosities, near IR/optical luminosities
(when the disc is visible),
and a standard Shakura \& Sunyaev (1973) disc emission model
is very powerful to find $M$ and $\dot M$.
With good data, showing the disc emission at its peak, the
uncertainty on the black hole mass is less than a factor 2.
Since the blazars here studied are $\gamma$--ray emitters,
we can also robustly constrain the jet power, since
the $\gamma$--ray luminosity, in these blazars,
is almost equal to the bolometric one.
We can then study in a robust way the link between the
accretion and the jet powers.

In \S 4.2 we discussed the uncertainties related to the jet power, 
associated to the unknown proton/lepton ratio.
One robust lower limit is associated to $P_{\rm r}$, the
power that the jet spends to produce the radiation we see.
It is simply $P_{\rm r}\sim \Gamma^2 L_{\rm bol}$ 
(see Appendix), where $L_{\rm bol}$ is the total jet luminosity.
Ghisellini \& Tavecchio (2010) have discussed the
importance of the Compton rocket effect on the jet when it is crossing
the BLR.
In the comoving frame of the jet, the seed photons are not isotropically distributed.
This implies that also the inverse Compton scattered photons
are not isotropic, and more power is emitted in the forward direction
(i.e. along the jet velocity direction).
The jet then must recoil.
If the jet is ``heavy" (i.e. one proton per electron) the
recoil is negligible, but if the jet is made by pairs the effect
is very important.
The jet halves its bulk Lorentz factor when there are $\sim$20 pairs
per proton.
This corresponds to a power that is roughly equal to $P_{\rm r}$.
Requiring that the jet {\it does not} decelerate significantly,
we end up with a minimum jet power (corresponding to a minimum
amount of protons) that simply is $P_{\rm j, min}=2P_{\rm r}$.
On the other hand, if we assume no pairs and therefore one proton
per electron, we have a jet power $P_{\rm j}$.

These two quantities are listed in the two last columns of Tab. \ref{powers}.
Fig. \ref{pjet} shows $P_{\rm j, min}$ (top panel) and $P_{\rm j}$ (bottom panel)
as a function of $L_{\rm d}$ for our blazars, where they are compared to other powerful
FSRQs that we have analysed in the past.
These are the FSRQs at $z>2$ detected by the 3 years all sky survey of {\it Swift}/BAT
(Ajello et al. 2009, these blazars are labelled BAT $z>2$ in Fig. \ref{pjet}
and were studied by Ghisellini et al. 2010a);
the FSRQs detected by {\it Fermi}/LAT in the first 3 months of operations
(labelled G10; Ghisellini et al. 2010b);
the 4 ``blue" quasars (FSRQs with a strong synchrotron component
peaking in the optical, labelled ``Blue QSOs" in Fig. \ref{pjet},
Ghisellini et al. 2012);
and all the FSRQs in the 1LAC sample present also in the
SDSS spectroscopic survey (labelled S12, Sbarrato et al. 2012a).

Note that:
\begin{itemize}

\item
Our blazars lie in the bulk of the distribution, 
with average values of the jet power and accretion luminosity.

\item
The correlation between $P_{\rm j, min}$ and $L_{\rm d}$ 
(top panel of Fig. \ref{pjet}) is significantly less dispersed 
than the $P_{\rm j}$--$L_{\rm d}$ relation.
We re--iterate that $P_{\rm j}$ is found considering one proton
per emitting electron, so that a non--constant number of pairs
per proton could be responsible for the larger dispersion.
However, we think it is premature to draw any strong
conclusion, given the related uncertainties.

\item
$P_{\rm j, min}$ is of the same order as $L_{\rm d}$.
Given that this jet power is a {\it lower} limit,
this suggest that the total jet power can be larger than $L_{\rm d}$.
In turn, this suggests that the origin of the jet power cannot be accretion only,
and favours the extraction of the black hole spin energy as the prime
movers of the jet power.

\item
We can compare the $z<2$ blazar detected by {\it Swift}/BAT with the
blazar in our sample. 
It is evident that the former have both more powerful jets 
and more luminous accretion discs, lying at the higher end of
the distribution of powers.

\end{itemize}

With our observational campaign all the {\it Fermi} $z>2$ blazars in the 2LAC catalog 
have been observed and detected in the {\it Swift}/XRT energy range.
This allows us to have a conclusive view of the SED of  high redshift
$\gamma$--ray blazars.

From our results we conclude that {\it Fermi} blazars at high redshifts are
indeed powerful, but not extreme.
Similarly, also their black hole masses are large, but not extreme.
This can be contrasted with high--$z$ blazars detected at hard X--ray energies,
that {\it all} have extreme values of the jet power, of the disc luminosity
and of the black hole mass.
Therefore the hard X--ray band ($>$30 keV) is more efficient than the $\gamma$--ray band
(given the current sensitivities) in finding the most powerful blazars.
This is expected if the so called ``blazar sequence" 
(Fossati et al. 1998; Ghisellini et al. 1998) holds
even at the highest power, since it predicts that the peak frequencies of
both the synchrotron and the high energy humps shift to lower values
when increasing the bolometric observed luminosity.
At the highest end of the power distribution, the high energy peak 
can shift to sub--MeV energies, implying a large hard X--ray flux
and a smaller $\gamma$--ray flux (with the K--correction working in the 
same direction).

%%%%%%%%%%%%%%%%%%%%%%%%%%%%%%%%%

\section*{Acknowledgements}
This research has made use of the NASA/IPAC Extragalactic Database (NED) which is 
operated by the Jet Propulsion Laboratory, California Institute of Technology, under 
contract with the National Aeronautics and Space Administration. 
Part of this work is based on archival data software or on--line services provided by the ASI
Data Center (ASDC).
This publication makes use of data products from the Wide-field Infrared Survey Explorer, 
which is a joint project of the University of California, Los Angeles, and the 
Jet Propulsion Laboratory/California Institute of Technology, funded by the 
National Aeronautics and Space Administration.
Part of the funding for GROND
(both hardware as well as personnel) was generously granted from the Leibniz
Prize to Prof. G. Hasinger (DFG grant HA 1850/28--1).

% This work was partially financed by a 2007 COFIN-- MiUR grant and
% by ASI grant I/088/06/0.

\vskip 2.5 cm
\noindent
{\bf APPENDIX}
\vskip 0.3 cm
\noindent
At a distance $R_{\rm diss}$ from the black hole of mass $M$
the jet dissipates
part of its power and injects relativistic electrons throughout the
emitting region, assumed to be spherical, with radius $R=\psi R_{\rm diss}$,
with $\psi=0.1$.  
In the region there is a tangled magnetic field $B$.
The relativistic electrons are injected with a smoothly joining broken power law
in energy: 
\begin{equation}
Q(\gamma)  \, = \, Q_0\, { (\gamma/\gamma_{\rm b})^{-s_1} \over 1+
(\gamma/\gamma_{\rm b})^{-s_1+s_2} } \quad {\rm [cm^{-3} s^{-1}]} 
\label{qgamma}
\end{equation}
The energy particle distribution $N(\gamma)$ [cm$^{-3}$] is calculated
solving the continuity equation where particle injection, radiative
cooling and pair production (via the $\gamma$--$\gamma \to e^\pm$
process), are taken into account. 
The created pairs contribute to the emission.

The injection process lasts for a light crossing time $R/c$, and we
calculate $N(\gamma)$ at this time.  This assumption comes from the
fact that even if injection lasted longer, adiabatic losses caused by
the expansion of the source (which is traveling while emitting) and
the corresponding decrease of the magnetic field would make the
observed flux to decrease.  Therefore the calculated spectra
correspond to the maximum of a flaring episode.

The total power injected into the source in the form of relativistic
electrons is $P^\prime_{\rm i}=m_{\rm e}c^2 V\int Q(\gamma)\gamma
d\gamma$, where $V=(4\pi/3)R^3$ is the volume of the emitting region.

The bolometric luminosity of the accretion disc
is $L_{\rm d}$.
Above and below the accretion disc, in its inner parts, there is an
X--ray emitting corona of luminosity $L_{\rm X}$ (it is fixed at a
level of 30\% of $L_{\rm d}$).  Its spectrum is a power law of energy
index $\alpha_X=1$ ending with a exponential cut at $E_{\rm c}=$150
keV.  The specific energy density (i.e. as a function of frequency) of
the disc and the corona are calculated in the comoving frame of the
emitting blob, and used to properly calculate the resulting External
inverse Compton spectrum.  
The BLR is assumed to be a thin spherical
shell, of radius $R_{\rm BLR}=10^{17} L_{\rm d, 45}^{1/2}$ cm.
We consider also the presence of a IR torus, at larger distances.  The
internally produced synchrotron emission is used to calculate the
synchrotron self Compton (SSC) flux.  Table \ref{para} lists the
adopted parameters.

The power carried by the jet can be in the form of
radiation ($P_{\rm r}$), magnetic field ($P_{\rm B}$), emitting
electrons ($P_{\rm e}$, no cold electron component is assumed) and
cold protons ($P_{\rm p}$, assuming one proton per emitting electron).
All the powers are calculated as
\begin{equation}
P_i  \, =\, \pi R^2 \Gamma^2\beta c \, U^\prime_i
\end{equation}
where $U^\prime_i$ is the energy density of the $i$ component, as
measured in the comoving frame.

The power carried in the form of the produced radiation, $P_{\rm r}
=\pi R^2 \Gamma^2\beta c \, U^\prime_{\rm rad}$, can be re--written as
[using $U^\prime_{\rm rad}=L^\prime/(4\pi R^2 c)$]:
\begin{equation}
P_{\rm r}  \, =\,  L^\prime {\Gamma^2 \over 4} \, =\, L {\Gamma^2 \over 4 \delta^4}
\, \sim \, L {1 \over 4 \delta^2}
\end{equation} 
where $L$ is the total observed non--thermal luminosity ($L^\prime$ is
in the comoving frame) and $U^\prime_{\rm rad}$ is the radiation
energy density produced by the jet (i.e.  excluding the external
components).  The last equality assumes $\theta_{\rm v}\sim 1/\Gamma$.

When calculating $P_{\rm e}$ (the jet power in bulk motion of emitting
electrons) we include their average energy, i.e.  $U^\prime_{\rm e}=
n_{\rm e} \langle\gamma\rangle m_{\rm e} c^2$.

\end{document}